\documentclass[prd,superscriptaddress,preprint,aps,amsmath,amssymb,showpacs,nofootinbib]{revtex4-2}

\usepackage{graphicx,enumerate,xcolor,enumitem,xspace,braket,mathrsfs} 
\usepackage[colorlinks=true,citecolor=blue]{hyperref}

\usepackage{comment}
\usepackage{nicefrac}
\usepackage{multirow}

\newcommand{\nn}{\nonumber\\}
\newcommand{\tr}{\operatorname{Tr}}

\newcommand{\tev}{\operatorname{TeV}}

\newcommand{\dk}{\operatorname{d\mathbf{k}}}
\newcommand{\dO}{\operatorname{d\Omega}}
\newcommand{\dphi}{\operatorname{d\phi}}

\newcommand{\hsigma}{\hat{\sigma}}
\newcommand{\sth}{s_\theta}
\newcommand{\cth}{c_\theta}

\newcommand{\suma}{ \frac{1}{N} \sum_{a =1}^N}
\newcommand{\intk}{\frac{1}{\sigma}\int \dO \frac{\rm d\sigma}{\dO}}
\newcommand\scalemath[2]{\scalebox{#1}{\mbox{\ensuremath{\displaystyle #2}}}}
\newcommand{\scrB}{\mathscr{B}}
\newcommand{\bfC}{C}
\newcommand{\ie}{{\it i.e.}}
\newcommand{\eg}{{\it e.g.}}

\newcommand{\Mt}{\mathcal{M}_t}
\newcommand{\Mu}{\mathcal{M}_u}
\newcommand{\Ms}{\mathcal{M}_s}
\newcommand{\pslash}{\!\not\! p}
\newcommand{\eslash}{\!\not\! \epsilon}

\NewCommandCopy{\oldvec}{\vec}
\renewcommand{\vec}[1]{\oldvec{\,#1}}

\usepackage[normalem]{ulem}

\newcommand{\Cijqqtt}{
    \begin{pmatrix}
        \frac{(2-\beta^2)\sth^2}{2-\beta^2\sth^2} & 0 & -\frac{2\cth\sth\sqrt{1-\beta^2}}{2-\beta^2\sth^2}\\
        0&\frac{-\beta^2\sth^2}{2-\beta^2\sth^2}&0\\
        -\frac{2\cth\sth\sqrt{1-\beta^2}}{2-\beta^2\sth^2} & 0 & \frac{2\cth^2 +\beta^2\sth^2}{2-\beta^2\sth^2}\\
    \end{pmatrix}
}
\newcommand{\CijggttSzero}{
    \begin{pmatrix}
        \frac{\beta^2-1}{\beta^2+1} & 0 & 0\\
        0&\frac{\beta^2-1}{\beta^2+1}&0\\
        0 & 0 & -1\\
    \end{pmatrix}
}
\newcommand{\Cijeedenomiator}[3]{#1^2(2-\beta^2\sth^2) + #2^2 \beta^2(1+\cth^2)  #3 4 #1 #2 \beta\cth }
\newcommand{\Cijeett}[3]{
\begin{pmatrix}
    \sth^2 \left(#1^2(2-\beta^2)- #2^2 \beta^2 \right) & 0 & -2 (#1^2\cth #3 #1#2\beta) \sth \sqrt{1-\beta^2}\\
    0&(#2^2-#1^2)\beta^2\sth^2&0\\
    -2 (#1^2\cth #3 #1#2\beta) \sth \sqrt{1-\beta^2} &0 & #1^2(2\cth^2+\beta^2\sth^2) + #2^2\beta^2(1+\cth^2) #3  4 #1#2\beta
\end{pmatrix}
}

\begin{document}

\title{Optimizing Entanglement and Bell Inequality Violation \\ 
in Top Anti-Top Events}
\author{Kun Cheng}
\email{kun.cheng@pitt.edu}
\affiliation{PITT PACC, Department of Physics and Astronomy,\\ University of Pittsburgh, 3941 O’Hara St., Pittsburgh, PA 15260, USA}
\affiliation{School of Physics, Peking University, Beijing, 100871, China} 

\author{Tao Han}
\email{than@pitt.edu}
\affiliation{PITT PACC, Department of Physics and Astronomy,\\ University of Pittsburgh, 3941 O’Hara St., Pittsburgh, PA 15260, USA}

\author{Matthew Low}
\email{mal431@pitt.edu}
\affiliation{PITT PACC, Department of Physics and Astronomy,\\ University of Pittsburgh, 3941 O’Hara St., Pittsburgh, PA 15260, USA}

\date{\today}

% \preprint{
% \begin{flushright}
% PITT-PACC-2401 
% \end{flushright}}
{\hfill PITT-PACC-2401}

\begin{abstract}
A top quark and an anti-top quark produced together at colliders have correlated spins.  These spins constitute a quantum state that can exhibit entanglement and violate Bell's inequality.   In realistic collider experiments, most analyses allow the axes, as well the Lorentz frame to vary event-by-event, thus introducing a dependence on the choice of event-dependent basis leading us to adopt ``fictitious states,'' rather than genuine quantum states.  The basis dependence of fictitious states allows for an optimization procedure, which makes the usage of fictitious states advantageous in measuring entanglement and Bell inequality violation.  In this work, we show analytically that the basis which diagonalizes the spin-spin correlations is optimal for maximizing spin correlations, entanglement, and Bell inequality violation.  We show that the optimal basis is approximately the same as the fixed beam basis (or the rotated beam basis) near the $t\bar t$ production threshold, while it approaches the helicity basis far above threshold.  Using this basis, we present the sensitivity for entanglement and Bell inequality violation in $t\bar t$ events at the LHC and a future $e^+e^-$ collider. Since observing Bell inequality violation appears to be quite challenging experimentally, and requires a large dataset in collider experiments, choosing the optimal basis is crucially important to observe Bell inequality violation.  Our method and general approach are equally applicable to other systems beyond $t \bar t$, including interactions beyond the Standard Model.
\end{abstract}

\maketitle
\newpage
{
\hypersetup{linkcolor=black}
\tableofcontents
}

%%%%%%%%%%%%%%%%%%%%%%%%%%%%%%%%%%%%%%%%%%%%%%%%%%%%%%%%%%%
\section{Introduction}

One of the most direct manifestations of quantum mechanics at high-energy colliders is the correlation between the spins of outgoing particles.  The $t\bar{t}$ system, where the spin of the top quark ($t$) and the spin of the anti-top quark ($\bar{t}$) are correlated~\cite{Barger:1988jj}, is one of the most studied and most experimentally promising systems. The spin of particles is not measured directly at high-energy colliders, but rather can be inferred by the statistical distributions of decay products~\cite{Mahlon:1995zn,Stelzer:1995gc}.  Moving beyond correlations between spins, a two-particle final state at colliders can be treated as a quantum state~\cite{Afik:2020onf} (see also~\cite{Fabbrichesi:2021npl,Severi:2021cnj,Afik:2022kwm,Aoude:2022imd,Aguilar-Saavedra:2022uye,Fabbrichesi:2022ovb,Afik:2022dgh,Aguilar-Saavedra:2022wam,Ashby-Pickering:2022umy,Severi:2022qjy,Altakach:2022ywa,Dong:2023xiw,Barr:2021zcp,Barr:2022wyq,Ashby-Pickering:2022umy,Aguilar-Saavedra:2022wam,Aguilar-Saavedra:2022mpg,Fabbrichesi:2023cev,Morales:2023gow,Fabbri:2023ncz,Bi:2023uop,Aguilar-Saavedra:2023hss,Aoude:2023hxv,Bernal:2023ruk,Ma:2023yvd,Sakurai:2023nsc,Han:2023fci,Cheng:2023qmz,Ehataht:2023zzt,Aguilar-Saavedra:2024fig,Maltoni:2024tul,Aguilar-Saavedra:2024hwd,Barr:2024djo,Aguilar-Saavedra:2024whi,Duch:2024pwm,Maltoni:2024csn,Afik:2024uif,Wu:2024asu}).  For such a quantum state, it would be very interesting to determine the entanglement between the top and anti-top and to measure whether Bell's inequality is violated. This would be a clear establishment of quantum information in high-energy collisions.

The magnitude of the spin correlation between outgoing particles depends on the choice of the quantization axis for the particle spin measurement. The simplest choice would be to use the $z$-axis of the beam direction as the quantization axis to measure the spin of the top, $S_z^t$, and the spin of the anti-top $S_z^{\bar{t}}$, the spin correlation is then
\begin{equation} \label{eq:spinCorrelation}
\mathscr{A} = \langle {S_z^t} \otimes S_z^{\bar{t}} \rangle.
\end{equation}
If we perform the measurement using a different basis the correlation would generally be different.  It has been customary to take the momentum direction of the top quark in the  $t\bar t$ center-of-mass frame as the spin quantization axis.  Using this basis would lead to a different correlation and therefore, the spin correlation as in Eq.~\eqref{eq:spinCorrelation} is basis-dependent.

The basis-dependence of spin correlations in the $t\bar{t}$ system is well understood for fixed initial states in hadronic collisions.  For the $q\bar{q}$ initial state, Mahlon and Parke calculated the maximal spin correlation to be 100\% and analytically determined the basis that achieves this~\cite{Mahlon:1995zn}.  For the $gg$ initial state, Uwer, Mahlon and Parke found that an optimal basis is also obtainable. In this case, however, the maximal spin correlation does not reach 100\% and varies over phase space \cite{Uwer:2004vp,Mahlon:2010gw}.

The entanglement of a quantum state, unlike the correlations between a pair of spins, is an intrinsic property of the state and does not depend on the basis choice of an observable. To treat the $t\bar{t}$ system as a quantum state, we first assume that a sufficient number of events are produced such that the $t\bar{t}$ phase space can be divided into infinitesimal bins.  The ideal way to measure the entanglement would be to only consider events from a single phase space bin.  Combining these events would result in a well-defined quantum state and its entanglement and Bell inequality violation would be basis-independent.

In order for repeated measurements to be usable to reconstruct a quantum state, we need an ensemble of \textit{identically-prepared quantum states}.  One example of identical preparation is only considering $t\bar t$ events from a fixed phase space point.  For the generalization to a region of phase space, as would be required in practice at colliders, identical preparation is still possible provided that the same quantization axis is used for all events in the phase space region.  This identically prepared spin state is obtained by projecting the underlying quantum state to a fixed phase space point, and we call this state a \textit{quantum sub-state}.

If measurements are performed with variable preparation, then the reconstruction does not lead to a quantum state.  Measuring the spin of the top along its momentum axis $-$ usually called the helicity basis $-$ where the top momentum axis points in a different direction in each event and corresponds to a different quantization axis for each event.  Rather than measuring a quantum state, one is instead measuring an angular-averaged sum of quantum sub-states.   These angular-averaged sums, when measured in an event-dependent basis, are not genuine quantum states.  They are a convex sum of quantum sub-states and are not added in a way that corresponds to a quantum state.  Ref.~\cite{Afik:2022kwm} labeled these as ``fictitious states'' and Ref.~\cite{Cheng:2023qmz} further elaborated their properties.

To see the basis-dependence of an angular-averaged sum, consider the density matrix $\bar{\rho}$ measured in an event-dependent basis
\begin{equation}
    \bar \rho = \sum_{a \in {\rm events}} \rho_a \  .
\end{equation}
Each $\rho_a$ is a quantum sub-state corresponding to a single event $a$.  To change basis, one rotates each sub-state by an event-dependent rotation $U_a$
\begin{equation}
\bar\rho
\to
\sum_{a \in {\rm events}} U^\dagger_{a} \rho_a U_{a} \ .
\end{equation}
Except in special cases, the event-dependent rotation does not correspond to an overall rotation $U$ of $\bar\rho$. In other words $\sum_{a \in {\rm events}} U^\dagger_{a} \rho_a U_{a} \neq U^\dagger \bar \rho U$.  Thus, when observables are computed from $\bar\rho$, the results depend on the basis that is used.

The special case where a common rotation is applied to all events corresponds to using a fixed basis. When reconstructing the spin density matrix using the same spin quantization axis in the center-of-mass frame of $t\bar t$ for all events, the angular-averaged sum is the projection of the underlying quantum state of $t\bar t$ into its spin space, and this results in a genuine quantum state. For a quantum state, the signals of entanglement and Bell inequality violation are basis-independent.
 
In practice, the majority of work on entanglement and Bell inequality violation at colliders utilizes the helicity basis, then the measurements are performed on fictitious states.  For a fictitious state, finding non-zero entanglement only shows that there exists a quantum sub-state that is entangled.  Likewise, finding a fictitious state that violates a Bell inequality only shows that there exists a quantum sub-state that violates a Bell inequality.  The usage of fictitious state, however, is still valuable to determine the existence of entanglement or Bell inequality violation in the $t\bar t$ system.

Since the size of entanglement is basis-dependent for fictitious states, there must exist an optimal basis that yields the largest entanglement and likewise for Bell inequality violation. In a previous publication \cite{Cheng:2023qmz}, we showed that the observation of the Bell inequality violation with the fictitious states will imply the same for the genuine bipartite qubit states. We analytically demonstrated that the basis that diagonalizes the spin-spin correlation is optimal for the construction of the fictitious states.  In this work, We extend the previous proofs in great detail in terms of the Lorentz properties and the collider observables. We show that the basis in which the $t\bar{t}$ spin correlation matrix is diagonal maximizes both the entanglement and the Bell inequality violation. We further present comparative results of the $t\bar t$ system for different basis choices at the CERN Large Hadron Collider (LHC) as well at an $e^+e^-$ collider. 

The $t$ and $\bar{t}$ (which we will also call a top pair) produced at the LHC are mainly entangled near threshold or in the high-$p_T$ region.  At sufficiently high energies, the optimal basis is close to the commonly used helicity basis. For example, at a large invariant mass of the system $m_{tt}\geq 1.2\ \tev$, the difference between the predicted violation of Bell's inequality between the helicity basis and the optimal basis is within 30\%.  Near threshold $m_{tt}\sim 2 m_t$, however, the optimal basis is closer to the beam basis~\cite{Uwer:2004vp}, while Bell inequality violation is nearly unobservable if we use the helicity basis.  Recently, the ATLAS~\cite{ATLAS:2023fsd} and CMS~\cite{CMS:2024pts,CMS-PAS-TOP-23-007} collaborations have observed quantum entanglement in $t\bar t$ system with a significance of $5\sigma$,  but whether Bell inequality violation can be measured with statistical significance, even with the full high luminosity LHC dataset, is unclear~\cite{Fabbrichesi:2021npl,Severi:2021cnj,Dong:2023xiw,Han:2023fci}.  Maximizing the sensitivity is therefore highly desirable.

Before we conclude this section, a special remark is in order regarding the generality of the test of the Bell inequality violation in collider experiments. References~\cite{Dreiner:1992gt,Abel:1992kz} pointed out that Bell inequality violation is not testable at a collider because quantities measured at colliders are commuting while non-commuting observables are required to violate a Bell inequality.  We would like to reiterate that in this work (and in other recent works on collider studies), we do not claim to perform a fully general ``test of quantum mechanics" against hidden variable theories.  Instead, within the framework of quantum field theory, we identify a quantum state to begin with in the collider environment, construct observables that measure the entanglement or Bell inequality violation, and then optimize these observables.  The non-commutation nature arises from the correlation information from the quantum density matrix and thus gains access to spins.  In a sense, we look to establish quantum tomography and quantum information in collider systems rather than question quantum mechanics itself.

The rest of this paper is arranged as follows. In Sec.~\ref{sec:quantumstates}, we reiterate the fact that properties of quantum states are independent of the quantization axis through explicit calculations. In Sec.~\ref{sec:fictitious}, we introduce the difference between genuine quantum states and fictitious states and elucidate the basis dependence of fictitious states.  In Sec.~\ref{sec:optimalbasis}, we prove that the basis that diagonalizes the correlation matrix, which is known to maximize spin correlations, also maximizes entanglement and Bell inequality violation.  In Sec.~\ref{sec:processes}, we give a detailed summary of the optimal basis in each Standard Model (SM) process that produces a top pair, and show how the optimal basis increases the signals of entanglement and Bell inequality violation in the $t\bar t$ system at the LHC and at a future $e^+e^-$ collider.  Finally, we conclude in Sec.~\ref{sec:conclusion}.  Some technical details for the basis transformation, entanglement of fictitious states, a general proof of the optimization basis, and the equivalence of the $gg\to t\bar t$ and $\gamma\gamma\to t\bar t$ scattering are provided in a few appendices.

%%%%%%%%%%%%%%%%%%%%%%%%%%%%%%%%%%%%%%%%%%%%%%%%%%%%%%%%%%%
\section{Quantum States in Different Bases and Frames}
\label{sec:quantumstates}

In this section, we elucidate properties of quantum states, namely that entanglement and Bell inequality violation do not depend on the basis choice of spin quantization nor on the Lorentz frame in which the spin is considered.

%%%%%%%%%%%%%%%%%%%%%%%%%%%%%%%%%%%%%%%%%%%%%%%%%%%%%%%%%%%
\subsection{Entanglement in Different Spin Bases}

Consider a bipartite qubit system. The state vector $\ket{\alpha\bar\alpha}\equiv \ket{\alpha}\otimes \ket{\bar\alpha}$ is a direct product of the spin state of each qubit. Its density operator $\hat \rho$ can be written as
\begin{equation}
\hat\rho=\rho_{\alpha\beta,\bar\alpha\bar\beta}\ket{\alpha\bar\alpha}\bra{\beta\bar\beta},
\quad\quad\quad
\alpha,\bar\alpha,\beta,\bar\beta
=1,2~(\text{or}~\uparrow,\downarrow).
\end{equation}
Here, the repeated spin indices are summed over 1 and 2 or over $\uparrow$ and $\downarrow$.  The states $\ket{\uparrow}$ and $\ket{\downarrow}$ are the positive and negative spin eigenstates along the third quantization axis $\vec e_3$, and the relative phase between $\ket{\uparrow}$ and $\ket{\downarrow}$ is determined by the first quantization axis $\vec e_1$.  Then $\rho_{\alpha\bar\alpha,\beta\bar\beta}$ is the $4\times 4$ matrix representation of $\hat \rho$ in the basis $\ket{\alpha\bar\alpha}$.

Next, consider choosing another spin basis $\ket{\alpha'}=\ket{\alpha}U_{\alpha\alpha'}$, which is equivalent to measuring the spin of each qubit along another quantization axis $\vec{e}_3'$. The new spin basis for the bipartite qubit system is  $\ket{\alpha'\bar\alpha'}=\ket{\alpha\bar\alpha}U_{\alpha\alpha',\bar\alpha\bar\alpha'}$ with $U_{\alpha\alpha',\bar\alpha\bar\alpha'}=U_{\alpha\alpha'}\otimes U_{\bar\alpha \bar\alpha'}$.  The operator $\hat\rho$ expressed in this new spin basis is $\hat \rho=\rho'_{\alpha'\beta',\bar\alpha'\bar\beta'}\ket{\alpha'\bar\alpha'}\bra{\beta'\bar\beta'}$.  The transformation of the density matrix from one basis to the other is given by
\begin{equation} \label{eq:UrhoU}
\rho'_{\beta\delta,\bar{\beta}\bar{\delta}}=U^\dagger_{\beta\alpha,\bar{\beta}\bar{\alpha}}\rho_{\alpha\gamma,\bar\alpha\bar{\gamma}}\,U_{\gamma\delta,\bar{\gamma}\bar{\delta}},
\end{equation}
or $\rho'=U^\dagger \rho\, U$ with the indices omitted.

Quantum entanglement, which is an intrinsic property of a quantum state, should not depend on the choice of basis for the spin measurement.  The concurrence is a criterion that provides a quantitative measurement of quantum entanglement. For a bipartite qubit system, the concurrence can be written as~\cite{Wootters:1997id}
\begin{equation}\label{eq:concurrence-lambda1234}
\mathscr{C}(\rho)=\max(0,\ \lambda_1-\lambda_2-\lambda_3-\lambda_4),
\end{equation}
where $\lambda_i$ ($i=1,2,3,4$) are the eigenvalues of $\sqrt{\sqrt{\rho}\tilde{\rho}\sqrt{\rho}}$ ordered in decreasing magnitude, $\tilde\rho=(\sigma_2\otimes \sigma_2)\rho^*(\sigma_2\otimes \sigma_2)$ with $\sigma_2$ the second Pauli matrix, and $\rho^*$ is the complex conjugate of $\rho$. A quantum state is entangled {\it if and only if} $~\mathscr C(\rho)>0$.  When we express the density matrix $\rho$ in a different basis as $\rho'$, we have $\rho'= U^\dagger \rho\, U$ and it follows that
\begin{align}\label{eq:sqrtrho}
\sqrt{\rho'} &= \sqrt{U^\dagger \rho\, U}= \sqrt{(U^\dagger \sqrt{\rho} \; U)(U^\dagger \sqrt{\rho} \; U)}=U^\dagger \sqrt{\rho} \; U,\\
\label{eq:tilderho}
\tilde{\rho}' &=(\sigma_2\otimes \sigma_2)U^T\rho^*U^* (\sigma_2\otimes \sigma_2)=U^\dagger (\sigma_2\otimes \sigma_2)\rho^* (\sigma_2\otimes \sigma_2) U= U^\dagger \tilde{\rho} \; U.
\end{align}
Therefore, the eigenvalues of $\sqrt{\sqrt{\rho}\tilde{\rho}\sqrt{\rho}}$ are invariant under the basis transformation, and the concurrence of a quantum state is basis-independent, \ie,
\begin{equation}
\mathscr{C}(\rho)=\mathscr{C}(U^\dagger\rho \; U).
\end{equation}

%%%%%%%%%%%%%%%%%%%%%%%%%%%%%%%%%%%%%%%%%%%%%%%%%%%%%%%%%%%
\subsection{Bell Inequality Violation in Different Spin Bases}

To discuss spin correlation and Bell inequality violation, it is convenient to decompose the density operator as
\begin{align}\label{eq:rhoBiCij}
\hat\rho=\frac{1}{4}\left( \hat{I}_2\otimes \hat{I}_2 + B_i^+ \hsigma_i\otimes \hat{I}_2 + B_i^- \hat{I}_2\otimes \hsigma_i + C_{ij} \hsigma_i\otimes \hsigma_j \right).
\end{align}
Here and after, repeated indices are summed from 1 to 3,  $\hat I_2=\ket{\uparrow}\bra{\uparrow}+\ket{\downarrow}\bra{\downarrow}$ is the identity operator on a qubit, and the spin operators $\hsigma_i=\ket{\alpha}\sigma_{i}^{\alpha\beta}\bra{\beta}$ are
\begin{align}
\hat\sigma_1=\ket{\uparrow}\bra{\downarrow}+\ket{\downarrow}\bra{\uparrow},
\quad\quad
\hat\sigma_2=-i\ket{\uparrow}\bra{\downarrow}+i\ket{\downarrow}\bra{\uparrow},
\quad\quad
\hat\sigma_3=\ket{\uparrow}\bra{\uparrow}-\ket{\downarrow}\bra{\downarrow},
\end{align}
where $\sigma_i$ are the Pauli matrices.
The density matrix $\hat\rho$ is parameterized by 15 real parameters, $B^\pm_i$ and $C_{ij}$, after choosing the spin operators.

When reparametrizing the density operator in another basis $\ket{\alpha'}=\ket{\alpha}U_{\alpha\alpha'}$, the spin operators are
\begin{align}
\hat\sigma_1'=\ket{\uparrow'}\bra{\downarrow'}+\ket{\downarrow'}\bra{\uparrow'},
\quad\quad
\hat\sigma_2'=-i\ket{\uparrow'}\bra{\downarrow'}+i\ket{\downarrow'}\bra{\uparrow'},
\quad\quad
\hat\sigma_3'=\ket{\uparrow'}\bra{\uparrow'}-\ket{\downarrow'}\bra{\downarrow'}.
\end{align}
The spin operators in two different bases satisfy the following transformation rule (see App.~\ref{sub:basisrotation} for details)
\begin{align}
\hat{\sigma}_i'= R_{ji} \hat{\sigma}_j,
\quad\quad\quad
R_{ij}=\frac{1}{2}\tr(U^\dagger \sigma_i U \sigma_j).
\end{align}
Therefore, in terms of the parameters $B_i^\pm$ and $C_{ij}$, the transformation rule of the density matrix in Eq.~\eqref{eq:UrhoU} can be  rewritten as
\begin{equation}
\label{eq:RCR}
B'^\pm_i
=R_{ij}^T B^\pm_j,
\quad\quad\quad
C'_{ij} =R_{ik}^TC_{k\ell} R_{\ell j}.
\end{equation} 
From the orthogonal condition of spin operators, we have that
\begin{equation}
\langle \hat \sigma_i \otimes \hat \sigma_j\rangle=C_{ij}, 
\quad\quad\quad
\langle \hat \sigma_i \otimes \hat I_2 \rangle=B_{i}^+, 
\quad\quad\quad
\langle \hat I_2\otimes \hat \sigma_i \rangle=B_{i}^-,
\end{equation}
where the expectation value of an operator $\hat{ \mathcal{O}}$ is $\braket{\hat{ \mathcal{O}}}\equiv\tr(\hat \rho \hat{\mathcal{O}})$.
Therefore, $B^\pm_i$ parameterizes the polarization of each particle and $C_{ij}$ parameterizes the spin correlation matrix.
The spin correlation, corresponding to an entry of the spin correlation matrix, is clearly basis-dependent from the above transformation rule. For a general unpolarized state, the maximum of the spin correlation, $\mathscr{A} = \langle S_z^t \otimes S_z^{\bar{t}} \rangle$, is achieved when the quantization axis $\vec{e}_3$ is an eigenvector of $C_{ij}$ with the largest eigenvalue~\cite{Uwer:2004vp}.\footnote{The spin correlation is taken as $\langle \sigma_i \otimes \sigma_j \rangle$ because this corresponds to a measurable differential cross section $d^2\sigma/d\ell_i d\bar{\ell}_j$, where $d\ell_i$ is an angle of one of the decay products of the top relative to the $i$th axis and $d\bar{\ell}_j$ is an angle of one of the decay products of the anti-top relative to the $j$th axis.  Using the component along the $z$-direction for both $t$ and $\bar{t}$ is conventional. In QCD, the $t\bar t$ production is $CP$-conserving and the correlation matrix $C_{ij}$ is symmetric so that the diagonal basis always exists.}  Following Ref.~\cite{Afik:2022kwm}, we will refer to the eigenbasis of $C_{ij}$ as the \textit{diagonal} basis.  Note that traditionally the diagonal basis for the $q\bar{q}\to t\bar t$ process is called the \textit{off-diagonal} basis because the amplitudes are only non-zero when the $t$ and $\bar{t}$ have unlike spins (\ie, off-diagonal in the space of spins)~\cite{Parke:1996pr,Mahlon:1997uc}. For general processes, a basis that makes the amplitudes off-diagonal is not guaranteed to exist.

Moving onto Bell inequalities, for local theories, the Clauser-Horne-Shimony-Holt (CHSH) inequality states that~\cite{Clauser:1969ny} 
\begin{align}\label{eq:Bella1a2b1b2}
\left| \Vec{a}_1\cdot C\cdot (\vec{b}_1-\Vec{b}_2)+
\Vec{a}_2\cdot C\cdot (\vec{b}_1+\Vec{b}_2) \right| \leq 2,
\end{align}
where $\vec a_1$, $\vec a_2$, $\vec b_1$, and $\vec b_2$ are four normalized spatial directions to measure the spins of $t$ and $\bar t$, and $C$ is the correlation matrix from Eq.~\eqref{eq:rhoBiCij}.  Classically this inequality is always satisfied while quantum theories may violate the inequality.  In this bipartite system, as formulated, the CHSH inequality can be violated.  The choice of the set of spatial directions determines the value of the left-hand side of Eq.~\eqref{eq:Bella1a2b1b2}.  Finding the largest violation of Bell's inequality corresponds to choosing the set of spatial directions that maximizes the left-hand side of Eq.~\eqref{eq:Bella1a2b1b2}.  We define the left-hand side of Eq.~\eqref{eq:Bella1a2b1b2} as the {\it ``Bell variable''} $\mathscr{B}(\rho)$ and the maximization corresponds to~\cite{HORODECKI1995340}
\begin{equation}\label{eq:Brho}
\mathscr{B}(\rho)=\max_{\vec a_1, \vec a_2, \vec b_1, \vec b_2}
\left| \Vec{a}_1\cdot C\cdot (\vec{b}_1-\Vec{b}_2)+
\Vec{a}_2\cdot C\cdot (\vec{b}_1+\Vec{b}_2) \right|
= 2\sqrt{\mu_1^2+\mu_2^2},
\end{equation}
where $\mu_1^2$ and $\mu_2^2$ are the largest two eigenvalues of $C^TC$ (if $C$ is symmetric, $\mu_{1,2}$ are the two largest magnitude eigenvalues of $C$.)  The eigenvalues of $C^TC$ are invariant under the basis transformation $C \to R^T C R$, therefore the violation of Bell's inequality is also basis independent
\begin{equation}
\mathscr{B}(\rho)=\mathscr{B}(U^\dagger\rho \, U).
\end{equation}

%%%%%%%%%%%%%%%%%%%%%%%%%%%%%%%%%%%%%%%%%%%%%%%%%%%%%%%%%%%
\subsection{Entanglement and Bell Inequality Violation in Different Lorentz Frames}
\label{sub:LorentzframeSpincorrelation}

In the case that top and anti-top are not simultaneously at rest, the entries of the spin correlation matrix measured in different Lorentz frames can be different.  This is because the spin of a spin-1/2 particle measured in two different Lorentz frames is related by a Wigner rotation which acts on a particle in its rest frame and the top and anti-top will have different Wigner rotations.

Consider a top quark that is in a momentum eigenstate.  The state vector is defined in both momentum space and spin space, so it can be written as 
\begin{equation}
\ket{t}=\ket{p_t}\otimes \ket{\phi}=\ket{p_t}\otimes \ket{\alpha}\phi_\alpha,
\end{equation}
where $\ket{p_t}$ denotes the momentum state with four-momentum $p_t$, $\ket{\phi}=\ket{\alpha}\phi_\alpha$ is the spin state in top quark rest frame, and $\alpha=\pm 1$ is the spin index.  The spin density matrix of concern is $\rho_{\alpha\alpha'}=\phi_\alpha \phi_{\alpha'}^*$.

The state vector observed in another Lorentz frame is
\begin{equation}
\ket{t}=\ket{p_t'}\otimes\ket{\phi'}=
\ket{p_t'}\otimes\ket{\alpha}\phi'_\alpha.
\end{equation}
The state vector observed in two different Lorentz frames satisfies the following relation~\cite{Gingrich:2002ota}
\begin{equation}
p_t'^\mu=\Lambda^\mu_\nu p_t^\nu, 
\quad\quad\quad
\phi'_{\alpha'} = U(\Lambda)_{\alpha'\alpha}^\dagger \phi_\alpha,
\end{equation}
where $\Lambda$ is the Lorentz transformation that connects these two frame choices, and the unitary matrix $U(\Lambda)$ is the spinor representation of a Wigner rotation that depends on $\Lambda$ and the momentum of the top quark; see Eq.~\eqref{eq:WignerR} for the explicit form.  Therefore, the relationship between the spin density matrix of a moving top quark observed in two different Lorentz frames is
\begin{equation}
\rho'_{\alpha'\beta'} = U(\Lambda)^\dagger_{\alpha'\alpha}  \rho_{\alpha\beta} U(\Lambda)_{\beta\beta'}.
\end{equation}
For the top pair system, both the top and the anti-top are moving but with different momenta.  Similarly, the density matrix of a quantum sub-state observed in one Lorentz frame is related to the density matrix in a different Lorentz frame by~\cite{Gingrich:2002ota}
\begin{equation}\label{eq:rhoLorentz}
\rho' = (U_t^\dagger \otimes U_{\bar t}^\dagger) \rho (U_t \otimes U_{\bar t}), 
\quad\quad\quad
U_t \neq U_{\bar t}.
\end{equation}
Here, the Wigner rotations for $t$ and $\bar t$ are different because their momenta are different.

From Eqs.~\eqref{eq:sqrtrho} and~\eqref{eq:tilderho}, we find that the concurrence of this state observed in two different Lorentz frames is still the same, \ie, $\mathscr{C}(\rho)=\mathscr{C}(\rho')$.  Combining Eq.~\eqref{eq:rhoLorentz} and Eq.~\eqref{eq:rhoBiCij}, we find that the spin correlation matrix measured in different Lorentz frames satisfies
\begin{equation}
C' = R_t^T C R_{\bar t},
\end{equation}
where $(R_t)_{ij}=\tr(U_t^\dagger \sigma_i U_t \sigma_j)/2$ and $(R_{\bar t})_{ij}=\tr(U_{\bar t}^\dagger \sigma_i U_{\bar t} \sigma_j)/2$.  Therefore, the eigenvalues of $C^T C$ are invariant under Lorentz transformations.  The size of Bell inequality violation, $\scrB(\rho)$, is similarly invariant under Lorentz transformations.

A special remark is that here we only consider quantum sub-states that are momentum eigenstates.  In the case where a state is entangled in spin and momentum space, the entanglement defined in the spin space is basis-dependent; see Ref.~\cite{Gingrich:2002ota} for details. 

Although the independence of the choice of frames and axes is illustrated in a bipartite qubit system, the basic properties relevant for entanglement and Bell inequality violation also hold in more general quantum systems.

\begin{figure}
  \centering
  \includegraphics[width=0.6\linewidth]{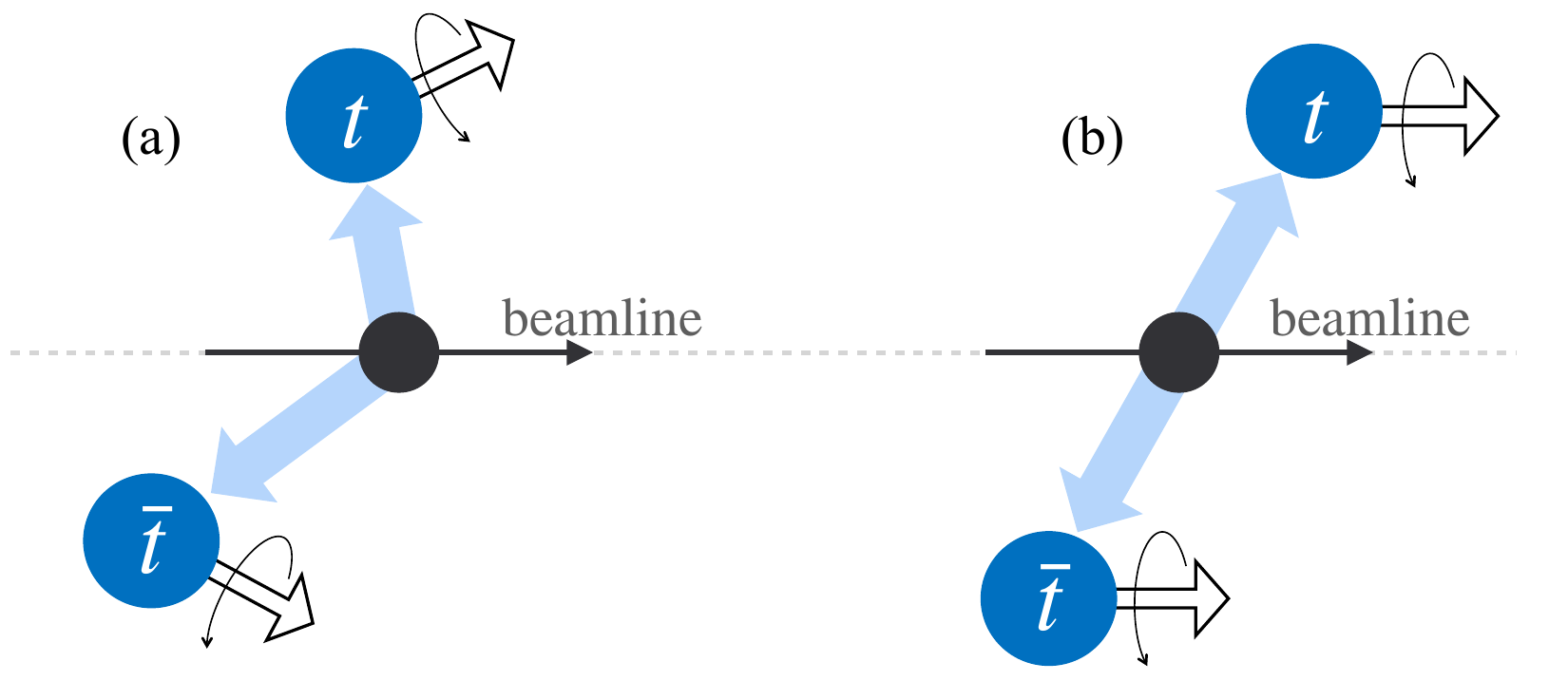}
  \caption{Illustration of the spins in a $t\bar t$ event produced at the LHC measured (a) in the lab frame and (b) the center-of-mass frame. The two frames are related by a Lorentz boost along the beamline direction.  The light blue arrows denote the moving directions of $t$ ($\bar t$) in the corresponding frame, and the hollow arrows represent the purposely chosen spin directions of $t$ ($\bar t$) in its rest frame.}
  \label{fig:PlotLorentzFrame}
\end{figure}

%%%%%%%%%%%%%%%%%%%%%%%%%%%%%%%%%%%%%%%%%%%%%%%%%%%%%%%%%%%
\section{Fictitious States}
\label{sec:fictitious}

The density matrix $\rho$ of a quantum state as discussed above is reconstructed from the expectation value of a series of measurements on a large event ensemble.  Denoting the quantum sub-state of each event as $\rho_a$, the reconstructed density matrix is an average of all $N$ quantum sub-states
\begin{equation}\label{eq:rhobarsuma}
    \bar \rho = \suma \rho_a.
\end{equation}
In collider experiments, each event, specified by the differential cross section at a multi-dimensional phase space point 
$d\sigma/d\Omega$, is a measurement of an underlying quantum state, and Eq.~\eqref{eq:rhobarsuma}, in the continuum limit, becomes
\begin{equation}\label{eq:rhobarintOmega}
\bar \rho = \intk \rho(\Omega).
\end{equation}
To obtain a genuine quantum state -- or a ``physical state'' -- from the average Eq.~\eqref{eq:rhobarsuma}, we need to write the density matrix $\rho_a$ in an event-independent basis, \ie, use the same axis $\hat e_i$ to measure the spin of all events $a$:
\begin{align}\label{eq:rhobarPhysical}
\bar\rho^{\vec e_i} &= \suma \rho_a^{\vec e_i}.
\end{align}
Here, $\rho_a^{\vec e_i}$ denotes the density matrix of the quantum sub-state of $a^{\rm th}$ event measured in the reference frame $(\vec e_1,\vec e_2,\vec e_3)$.  Note that all the physical states are equivalent, \ie, the physical state averaged in two different bases are equal with respect to a unitary rotation, $\bar \rho^{\vec e_i} = U^\dagger \bar\rho^{\vec e'_i} U$.

When measuring the spin of each qubit in an event $a$ in a different basis $\vec{e}_{i,a}$, however, the average in Eq.~\eqref{eq:rhobarsuma} gives 
\begin{equation}\label{eq:rhobarFictitious}
\bar\rho[\vec e_{i,a}] = \suma \rho_a^{\vec e_{i,a}}
= \suma U^\dagger_{a} \rho_a^{\vec e_{i}} U_{a}.
\end{equation}
A state averaged in an event-dependent basis is generally not equivalent to the physical state, \ie, there is no guarantee of a unitary transformation $U$, such that $\bar \rho[\vec e_{i,a}]=U^\dagger \bar \rho^{\vec e_i} U$.  Therefore, the density matrix averaged in an event-dependent basis is a ``fictitious state,''  rather than a genuine quantum state.  Moreover, the density matrix averaged in two different event-dependent bases $\vec e_{i,a}$ and $\vec e_{i,a}'$ are not equivalent, \ie, there is no guarantee of a unitary transformation $U$, such that $\bar \rho[\hat e_{i,a}]=U^\dagger \bar \rho[\hat e'_{i,a}] U$.  In other words, the fictitious states are basis-dependent.

We can define a fictitious state as the result of constructing a state according to Eq.~\eqref{eq:rhobarsuma} but using an event-dependent basis.  As shown above, this is not a quantum state because it does not retain all of the properties that a quantum state has.  We still call it a state (rather than simply a sum), however, because it does retain some properties of quantum states.  One example is the existence of Bell inequality violation where a fictitious state that has Bell inequality violation implies the {\it existence} of a quantum sub-state that has Bell inequality violation.

The usage of event-dependent bases and fictitious states may seem unnatural for low-energy experiments. At high-energy experiments, however, both event-dependent spin axis and Lorentz frame choices are conventionally used.  We next show that either event-dependent spin axis or event-dependent Lorentz frame choices lead to fictitious states.

%%%%%%%%%%%%%%%%%%%%%%%%%%%%%%%%%%%%%%%%%%%%%%%%%%%%%%%%%%%
\subsection{Fictitious States From Angular-Averaged Sums}

Experimentally, the polarization of top quarks cannot be measured directly, so the spin density matrix is reconstructed from the angular distributions of the final state particles.  For example, in the di-lepton decay mode, the correlation matrix is
\begin{equation}\label{eq:Cijlilj}
C_{ij}=-9\braket{ \cos\theta^+_i \cos\theta^-_j }, 
\quad\quad\quad
\cos\theta^\pm_i = \vec{p}_{\ell^\pm} \cdot \vec{e}_i,
\end{equation}
where $\vec p_{\ell^\pm}$ is the normalized three-momentum of the lepton $\ell^+$ ($\ell^-$) in the rest frame of the $t$ ($\bar t$), and $\vec e_i$ is the quantization axis used to determine the lepton momentum direction. Ideally, with an infinite number of events, one can study the $t\bar t$ system produced at a fixed phase space point $\Omega$, so that all the events are measured with the same quantum sub-states, determined by $C_{ij}(\Omega)$. In reality, each measured event corresponds to a top pair produced at a different phase space point.  Therefore, with Eq.~\eqref{eq:Cijlilj}, one obtains the correlation matrix of an angular-averaged sum $\bar \rho$ in Eq.~\eqref{eq:rhobarintOmega}:\footnote{As a reminder, we use angular-averaged sum as the general description which may correspond to either a quantum state or a fictitious state.}
\begin{equation}\label{eq:CbarPhysical}
    \bar C = \intk C(\Omega).
\end{equation}
As discussed in Ref.~\cite{Cheng:2023qmz}, to obtain a physical quantum state we must choose a set of $\vec e_i$ that are independent of the top quark momentum with respect to the average in Eq.~\eqref{eq:CbarPhysical}.  

In principle, the lab frame in any collision is a genuine event-independent frame convenient for measurements and analyses. The center-of-mass frame of $t\bar t$ system, however, is the most commonly used frame that captures the underlying dynamics when studying $t\bar t$ production at hadron colliders.  As illustrated in Fig.~\ref{fig:PlotLorentzFrame}, the lab frame in hadronic collisions and the center-of-mass frame of the $t\bar t$ system are, in general, related by a Lorentz boost along the beam direction.  The center-of-mass frame is an event-dependent frame, governed by the interaction and kinematics of the initial state partons. 

\begin{figure} 
  \centering
  \includegraphics[width=.4\linewidth]{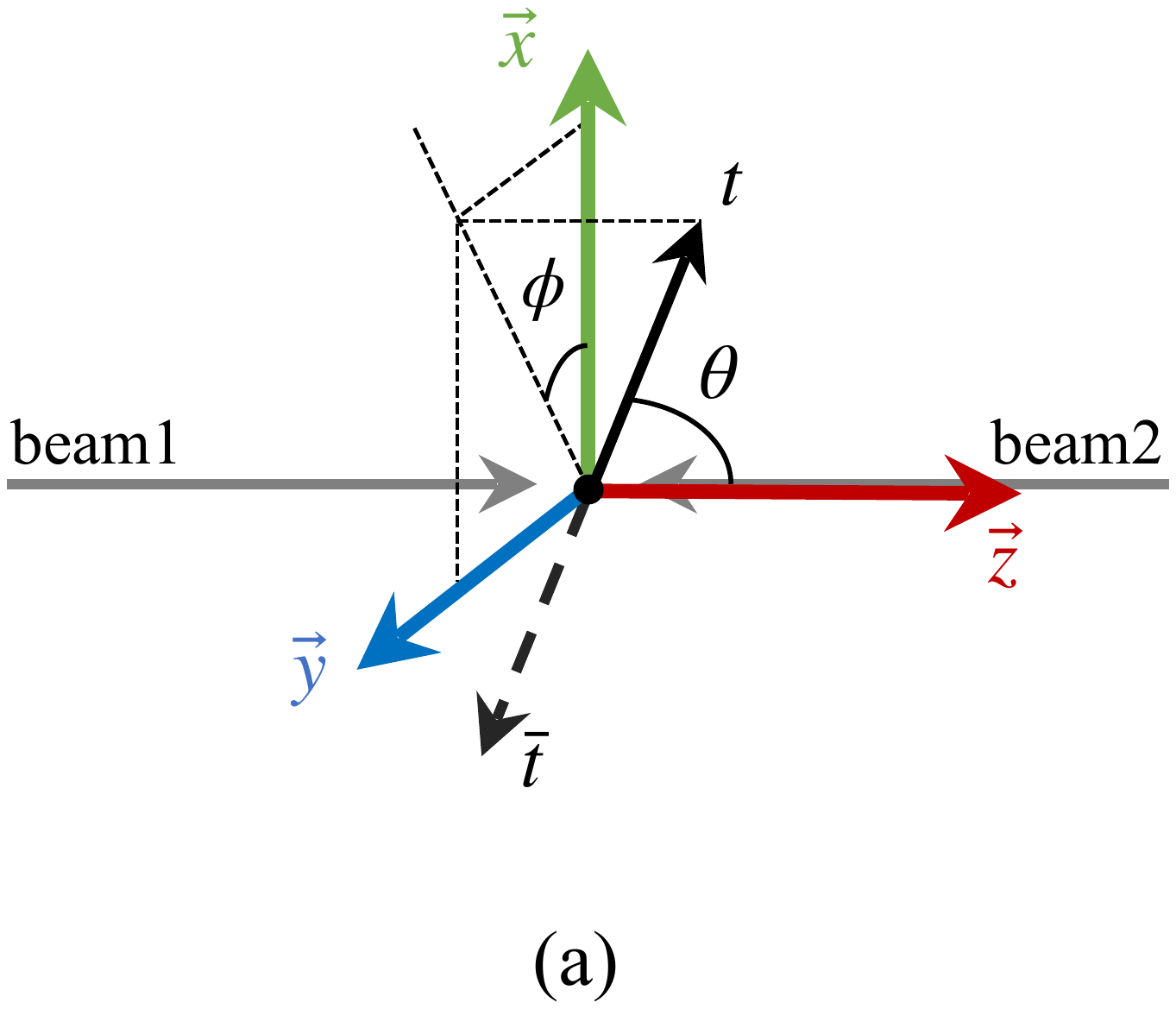}
  \includegraphics[width=.4\linewidth]{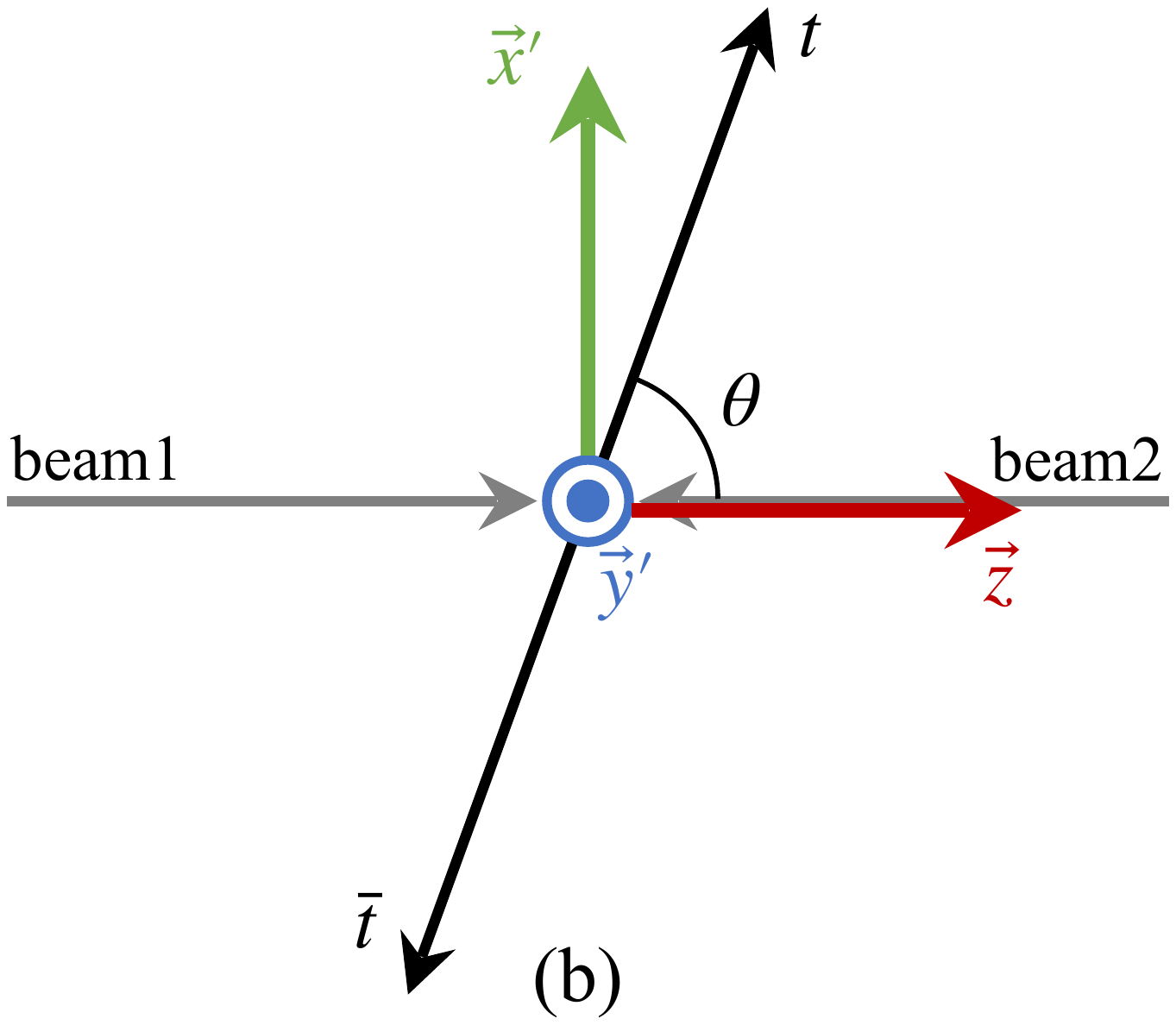}
  \includegraphics[width=.4\linewidth]{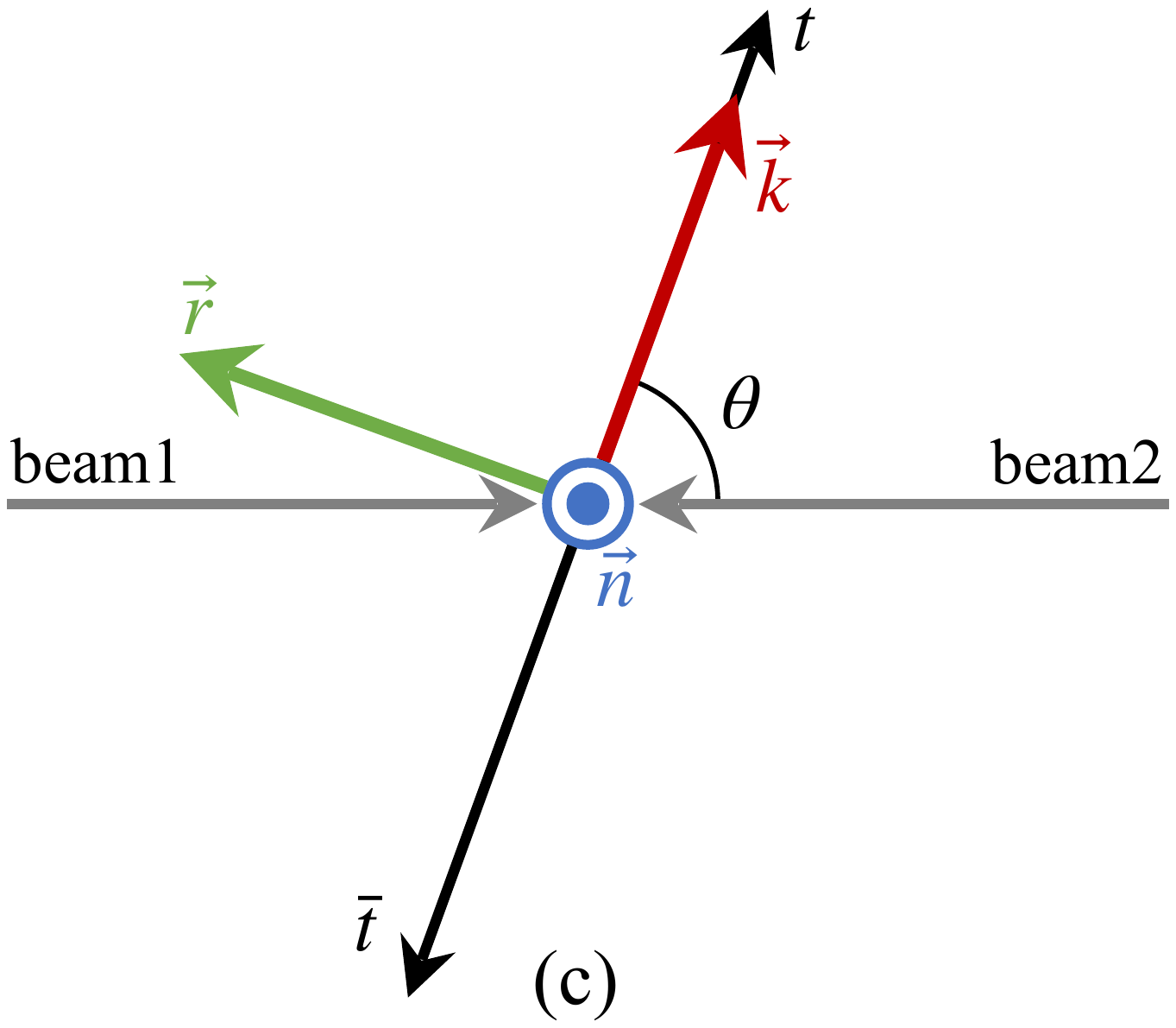}
  \includegraphics[width=.4\linewidth]{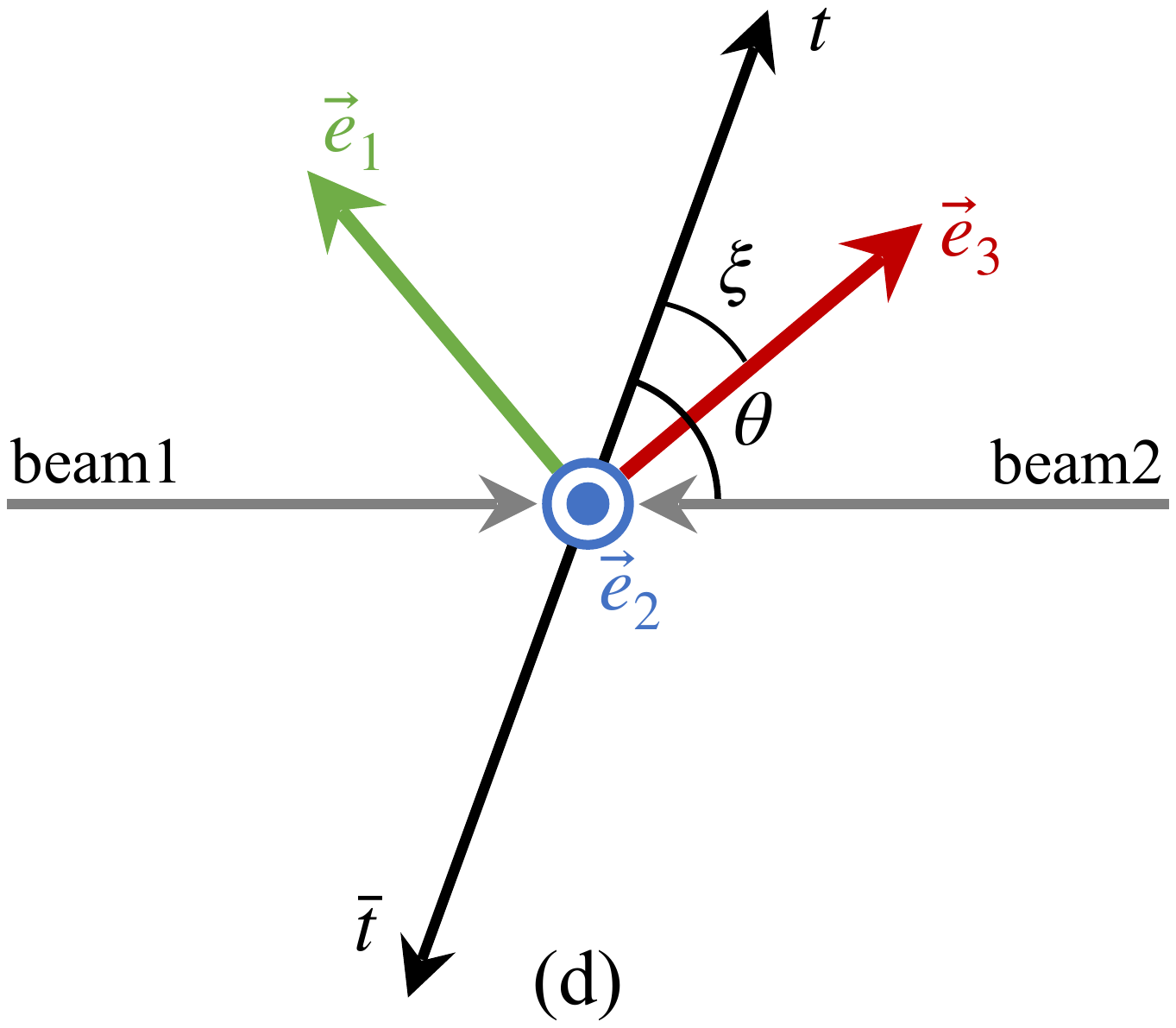}
  \caption{(a) Fixed beam basis $(\vec x,\vec y,\vec z)$. (b) Rotated beam basis $(\vec x',\vec y',\vec z)$. (c) Helicity basis $(\vec r,\vec n,\vec k)$. (d) Diagonal basis $(\vec e_1^{\,\rm diag},\vec e_2^{\,\rm diag},\vec e_3^{\,\rm diag})$. The rotation angle $\xi$ of the diagonal basis depends on the production process, \eg, $\tan\xi=(1/\gamma)\tan\theta$ for $q\bar{q}\to t\bar{t}$~\cite{Mahlon:1997uc}. }
  \label{fig:PlotCoordinates}
\end{figure}

Examples of event-dependent bases are shown in Fig.~\ref{fig:PlotCoordinates}, where we have the \textit{fixed beam basis} in Fig.~\ref{fig:PlotCoordinates}(a) with an arbitrary azimuthal angle orientation, the \textit{rotated beam basis} in Fig.~\ref{fig:PlotCoordinates}(b) in line with the vectors $\vec p_t$ and $\vec z$ forming the scattering plane, the \textit{helicity basis} in Fig.~\ref{fig:PlotCoordinates}(c) with the spin along the $\vec p_t$ direction, and the \textit{diagonal basis} in Fig.~\ref{fig:PlotCoordinates}(d) to be worked out later.\footnote{To construct the rotated beam basis and the helicity basis, only knowledge of the event-by-event kinematics is required, whereas the diagonal basis also requires knowledge of the production density matrix. At leading order, in the helicity basis the correlation matrix $C_{ij}$ only has one off-diagonal entry $C_{kr}$ such that only one rotational angle $\xi$ in the $\vec{k}$-$\vec{r}$ plane is needed to determine the diagonal basis.  This angle is shown in Fig.~\ref{fig:PlotCoordinates}.}  Depending on the specific physical process, a certain basis may be advantageous. For instance, when the invariant mass of the $t\bar{t}$ system is large, the helicity basis is most commonly used. 

In practice, event-dependent bases are commonly used when averaging the lepton momentum. The basis $\vec e_i$ in Eq.~\eqref{eq:Cijlilj} is replaced with an event-dependent basis $\vec e_i(\Omega)$, and the average in Eq.~\eqref{eq:Cijlilj} yields
\begin{equation}\label{eq:CbarFictitious}
    \bar C^{\rm fic} = \intk R^T(\Omega) C(\Omega) R(\Omega),
\end{equation}
where $R(\Omega)$ is the event-dependent rotation that rotates the fixed basis to $\vec e_i(\Omega)$,  and $\bar C^{\rm fic}$ is the correlation matrix of the fictitious state constructed in Eq.~\eqref{eq:rhobarFictitious}.

%%%%%%%%%%%%%%%%%%%%%%%%%%%%%%%%%%%%%%%%%%%%%%
\subsection{Fictitious States From Event-Dependent Lorentz Frame Choices}

In addition to event-dependent spin quantization axis choices, an event-dependent Lorentz frame choice also results in a fictitious state, such as a bipartite $t\bar t$ state constructed in the center-of-mass frame, in contrast to that in the lab frame. 
We start from one event in which the four-momentum of top, anti-top, and their decayed leptons in the lab frame are $p_t^{\rm lab}$, $p_{\bar t}^{\rm lab}$, $p_{\ell^+}^{\rm lab}$, and $p_{\ell^-}^{\rm lab}$, respectively.  When reconstructing the top pair spin correlation matrix in the rest frames of the top and the anti-top, we boost the lepton momenta in the lab frame to the rest frames of top and anti-top quark, respectively, 
\begin{align}\label{eq:plinlabframe}
p_{\ell^+} &= L(p_t^{\rm lab}) p_{\ell^+}^{\rm lab}, 
\quad\quad\quad
p_{\ell^-} = L(p_{\bar t}^{\rm lab}) p_{\ell^-}^{\rm lab},
\end{align}
where $L(p)$ is a pure Lorentz boost from the lab frame to the rest frame of the particle with momentum $p$.  Next, consider reconstructing the correlation matrix in another (event-dependent) frame, where the four momentum of each particle are $p_t^{\Lambda}$, $p_{\bar t}^{\Lambda}$ $p_{\ell^+}^{\Lambda}$, and $p_{\ell^-}^{\Lambda}$.  This frame is related to the lab frame by a Lorentz transformation $\Lambda$, so that $p^\Lambda = \Lambda p^{\rm lab}$.  Boosting the lepton momenta measured in this frame to the rest frames of the $t$ and the $\bar t$ yields
\begin{align}\label{eq:plinanotherframe}
p_{\ell^+}' &=L(p_t^{\Lambda}) p_{\ell^+}^{\Lambda}, 
\quad\quad\quad
%p_{\ell^-}' =L(p_t^{\Lambda}) p_{\ell^+}^{\Lambda}.
p_{\ell^-}' =L(p_{\bar{t}}^{\Lambda}) p_{\ell^-}^{\Lambda}.
\end{align}
Combining Eqs.~\eqref{eq:plinlabframe} and~\eqref{eq:plinanotherframe}, we find that $p_\ell$ and $p_\ell'$ are related by a Lorentz transformation
\begin{align}
p_{\ell^+}' &=L(p_t^{\Lambda}) p_{\ell^+}^{\Lambda} 
=L(\Lambda p_t^{\rm lab}) \Lambda p_{\ell^+}^{\rm lab} 
=L(\Lambda p_t^{\rm lab}) \Lambda  L^{-1}(p_t^{\rm lab}) \; p_{\ell^+}, \nn
% p_{\ell^-}' &=L(p_t^{\Lambda}) p_{\ell^-}^{\Lambda} 
% =  L(\Lambda p_{\bar t}^{\rm lab})\Lambda  p_{\ell^-}^{\rm lab} 
% =  L(\Lambda p_{\bar t}^{\rm lab})\Lambda L^{-1}(p_{\bar t}^{\rm lab}) \; p_{\ell^-}.
p_{\ell^-}' &=L(p_{\bar{t}}^{\Lambda}) p_{\ell^-}^{\Lambda} 
=  L(\Lambda p_{\bar t}^{\rm lab})\Lambda  p_{\ell^-}^{\rm lab} 
=  L(\Lambda p_{\bar t}^{\rm lab})\Lambda L^{-1}(p_{\bar t}^{\rm lab}) \; p_{\ell^-}.
\end{align}
The Lorentz transformation $L(\Lambda p) \Lambda L^{-1}(p)$ is a pure rotation, called a Wigner rotation
\begin{equation}\label{eq:WignerR}
L(\Lambda p) \Lambda L^{-1}(p) = 
\begin{pmatrix}
    1 & 0 \\
    0 & R(\Lambda, p)
\end{pmatrix}.
\end{equation}
Consequently, the reconstructed lepton momenta with these two different methods are related by the Wigner rotation,
\begin{equation}
\vec{p}_{\ell^+}' =  R(\Lambda, p_{t}^{\rm lab})  \vec{p}_{\ell^+}, 
\quad\quad\quad
\vec{p}_{\ell^-}' =  R(\Lambda, p_{\bar t}^{\rm lab})  \vec{p}_{\ell^-}.
\end{equation}
Here, these two $3\times 3$ rotations depend on the four-momentum of top or anti-top quark and the Lorentz transformation $\Lambda$ between two different frames.  We abbreviate the Wigner rotation $R(\Lambda,p_t^{\rm lab})$ and $R(\Lambda,p_{\bar t}^{\rm lab})$ as $R_t(\Lambda)$ and $R_{\bar t}(\Lambda)$ in the following.

Therefore, as introduced in Sec.~\ref{sub:LorentzframeSpincorrelation}, the correlation matrix of a quantum sub-state measured in two different Lorentz frames satisfies the relation
\begin{equation}
    C'(\Omega) = R_t^T(\Lambda_{\Omega}) C(\Omega) R_{\bar t}(\Lambda_{\Omega}).
\end{equation}
We find that measuring the correlation matrix in a different Lorentz frame is equivalent to measuring the correlation matrix in the original Lorentz frame but with a different spin axis of $\vec{\, e}_{i,\Omega}^{\,t}$ and $\vec{\, e}_{i,\Omega}^{\,\bar t}$ for $t$ and $\bar t$, respectively, where $ \vec{\, e}_{i,\Omega}^{\,t} = R_{t,ji}(\Lambda_\Omega) \vec{e}_j $ and $ \vec{\, e}_{i,\Omega}^{\,\bar t} = R_{\bar{t},ji}(\Lambda_\Omega) \vec{e}_j $.   Here $\vec e_j$ is the spin axis used in the original Lorentz frame.

Thus our formalism of fictitious states in Eq.~\eqref{eq:rhobarFictitious} applies equally for event-dependent Lorentz frame choices.  When reconstructing Eq.~\eqref{eq:Cijlilj} with event-dependent Lorentz frames, the resulting correlation matrix is
\begin{align}
\bar C^{\rm fic} = \intk R_t(\Lambda_\Omega) C(\Omega) R_{\bar t}(\Lambda_\Omega)^T,  
\end{align}
where $\Lambda_\Omega$ denotes the Lorentz transformation from the lab frame to an event-dependent Lorentz frame.  Similar to Eq.~\eqref{eq:CbarFictitious}, the averaged correlation matrix $\bar C^{\rm fic}$ also depends on the event-by-event rotations $R_t(\Lambda_\Omega)$ and $R_{\bar t}(\Lambda_\Omega)$.

%%%%%%%%%%%%%%%%%%%%%%%%%%%%%%%%%%%%%%%%%%%%%%%%%%%%%%%%%%
\subsection{Basis Dependence of Spin Correlation and Entanglement}

Due to the choice of event-dependent bases, fictitious states are conventionally utilized when discussing quantum observables of a ``state'' produced at colliders.  While the entanglement and Bell inequality violation of quantum states are basis-independent as we demonstrated earlier, observables computed using fictitious states and fictitious states themselves are basis-dependent in general.  For example, the entanglement and Bell inequality violation of a fictitious state becomes basis-dependent. Consequently, this gives us the freedom to find a basis that optimizes the measurement of entanglement and Bell inequality violation.

Before digging into details of optimizing the entanglement and Bell inequality violation, it is worth pointing out that the basis dependence of the entanglement of fictitious states is different from the basis dependence of spin correlations, which has been thoroughly studied for the $t\bar t$ system~\cite{Parke:1996pr,Mahlon:1997uc,Uwer:2004vp}.

The spin correlation only depends on the quantization axis chosen to measure the spin, \eg, defining $\ket{\uparrow}$ and $\ket{\downarrow}$ with respect to the third quantization axis $\vec e_3$.  The size of spin correlation does not depend on the phases of $\ket{\uparrow}$ and $\ket{\downarrow}$.  On the other hand, angular-averaged sums do depend on these phases. Equivalently, they depend on all three quantization axes $\vec e_i(\mathbf{k})$. For example,  in the high-$p_T$ region, a maximally-entangled Bell triplet state is produced from both the $q\bar q$ and the $gg$ initial states 
\begin{equation}
\ket{\Psi(\phi)} =  \frac{e^{-i\phi}\ket{\uparrow\uparrow}+e^{i\phi}\ket{\downarrow\downarrow}}{\sqrt{2}}.
\end{equation}
Here, the relative phase of the state is calculated in fixed beam basis.  The azimuthal angular average of the density matrix in the fixed beam basis is
\begin{equation}
  \bar \rho^{\rm fixed}_{\rm triplet}
=\frac{1}{2\pi} \int \dphi \ket{\Psi(\phi)}\bra{\Psi(\phi)} =\frac{\ket{\uparrow\uparrow}\bra{\uparrow\uparrow}+\ket{\downarrow\downarrow}\bra{\downarrow\downarrow}}{2},
\end{equation}
As the azimuthal angle in the rotated beam basis is always chosen to be zero, the azimuthal angular average of the density matrix in rotated beam basis is
\begin{align}
   \bar \rho^{\rm rotated}_{\rm triplet}&= \frac{1}{2\pi} \int\dphi \ket{\Psi(0)}\bra{\Psi(0)}=\ket{\Psi(0)}\bra{\Psi(0)},
\end{align}
where $\ket{\Psi(0)}=\ket{\Psi(\phi=0)} = (\ket{\uparrow\uparrow}+\ket{\downarrow\downarrow})/\sqrt{2}$.  The angular-averaged sum in the fixed basis $\bar\rho^{\rm fixed}_{\rm triplet}$ is a separable state, while $\bar\rho^{\rm rotated}_{\rm triplet}$ is still a maximally entangled state, although the spin correlation $\langle S_3^t \otimes S_3^{\bar t} \rangle=1$ is the same for both angular-averaged sums.

%%%%%%%%%%%%%%%%%%%%%%%%%%%%%%%%%%%%%%%%%%%%%%%%%%%%%%%%%%%
\section{The Optimal Basis for Entanglement and Bell Inequality Violation}
\label{sec:optimalbasis}

In this section, we prove that the diagonal basis maximizes the entanglement and the Bell inequality violation for fictitious states.   For simplicity, we first present our proof assuming the spin correlation matrix is symmetric, which is true in the center-of-mass frame of the $t\bar t$ system when the $t\bar t$ events are produced from $CP$-conserving processes (as in the SM).

We start with an arbitrary event-dependent basis choice $\vec e_{i,\Omega}$.  The correlation matrix in this basis is
\begin{equation}\label{eq:Cbardiagonal}
\bar \bfC_{ij} =\intk \bfC_{ij,\Omega}, 
\quad\quad\quad
{\rm with}\ 
(\bar c_1, \bar c_2, \bar c_3)\ {\rm as}\ {\rm eigenvalues\ of}\  \left[\bar \bfC \right].
\end{equation}
Here, we omit the basis choice label for simplicity when referring to a general event-dependent basis.

When the correlation matrix for each sub-state is symmetric, it can be diagonalized with an event-dependent rotation:\footnote{ Without loss of generality, the eigenvalues are ordered as $\mu_{1,\Omega} \geq\mu_{2,\Omega} \geq \mu_{3,\Omega} $. }
\begin{equation}\label{eq:Cdiagonal}
\bfC^{\rm diag}_\Omega=R_{\Omega}\bfC_\Omega R^T_{\Omega}=\begin{pmatrix}
        \mu_{1,\Omega} &0&0\\
        0&\mu_{2,\Omega} &0\\
        0&0&\mu_{3,\Omega} \\
    \end{pmatrix}.
\end{equation}
This diagonal spin correlation matrix is obtained if one measures the spin of $t$ and $\bar t$ with the following quantization axis
\begin{equation}
    \vec{\, e}_{i,\Omega}^{\,\rm diag} = R_{ij} \vec{e}_j,
\end{equation}
where $\vec{e}_{i,\Omega}^{\,\rm diag}$ are three eigenvectors of $\bfC_{\Omega}$ and they are referred to as the diagonal basis.

In the diagonal basis, the eigenvalues of the angular-averaged spin correlation matrix are
\begin{equation}\label{eq:mubardefinition}
\bar c^{\rm diag}_i = \intk \mu_{i,\Omega} \equiv \bar \mu_i.
\end{equation}
In any other basis, the eigenvalues of angular-averaged spin correlation matrix $\bar C$, namely $\bar c_i$, are  generally not equal to the average of $\mu_i$, but they satisfy the following relations~\cite{Cheng:2023qmz},
\begin{align}
\label{eq:trC1trC2}
&\bar c_1 + \bar c_2 + \bar c_3=\bar \mu_1 + \bar \mu_2 + \bar \mu_3 = \tr(\bar C), \\ 
\label{eq:mu1Ciimu3}
&\bar \mu_1 \geq \bar c_{i} \geq \bar \mu_3.
\end{align}
Eq.~\eqref{eq:trC1trC2} is a direct consequence of taking the trace of Eqs.~\eqref{eq:Cbardiagonal} and~\eqref{eq:Cdiagonal}.  To prove Eq.~\eqref{eq:mu1Ciimu3}, which states that $\bar c_i$ are bounded by $\bar \mu_i$, we first denote the three eigenvectors of $\overline \bfC$ as $\hat{v}_i$.  The corresponding eigenvalue is then
\begin{align}
    \bar c_i=\hat{v}_i\cdot \overline \bfC \cdot \hat{v}_i
    =\intk \left( \hat{v}_i\cdot \bfC_\Omega \cdot \hat{v}_i \right).
\end{align}
Applying Eq.~\eqref{eq:Cdiagonal} and $\bfC_\Omega=R_\Omega \bfC^{\rm diag}_\Omega R_\Omega^T$, we find that 
\begin{equation}
\hat{v}_i\cdot \bfC_\Omega \cdot \hat{v}_i=\sum_{\ell=1}^3 \left|(R_\Omega \cdot \hat v_i)_\ell\right|^2 \mu_{\ell,\Omega},
\end{equation}
is a convex sum of $\mu_{i,\Omega}$.  This leads to
\begin{equation}
\label{eq:mu1kCkmu3k}
    \mu_{1,\Omega} \geq \hat{v}_i\cdot \bfC_\Omega \cdot \hat{v}_i \geq \mu_{3,\Omega}.
\end{equation}
Therefore, Eq.~\eqref{eq:mu1Ciimu3} holds as a convex sum of Eq.~\eqref{eq:mu1kCkmu3k}.

With Eqs.~\eqref{eq:trC1trC2} and~\eqref{eq:mu1Ciimu3}, we are ready to prove that the diagonal basis maximizes both $\mathscr{B}(\bar\rho)$ and $\mathscr{C}(\bar\rho)$.

%%%%%%%%%%%%%%%%%%%%%%%%%%%%%%%%%%%%%%%%%%%%%%%%%%%%%%%%%%%
\subsection{Concurrence}
\label{sub:concurrence}

We focus on the unpolarized $t\bar t$ system at leading order where the polarization vectors $B_i^\pm$ are zero.  In this case $\tilde\rho = \rho$, as can be seen from the construction below Eq.~\eqref{eq:concurrence-lambda1234}.   After some explicit calculation~\cite{Afik:2020onf,Afik:2022kwm}, the concurrence is expressed as follows, with the largest eigenvalue labelled $\lambda(\rho)$:
\begin{equation}
    \mathscr{C}(\rho)=\max(2\lambda(\rho)-1,0).
\end{equation}
To compare the concurrence of the angular-averaged density matrix in different bases, we only need to compare their largest eigenvalues.  Substituting the correlation matrix into Eq.~\eqref{eq:rhoBiCij}, we find the four eigenvalues of an angular-averaged density matrix, $\bar\rho$, are
\begin{equation}
\frac{1-\tr(\bar C)}{4},\quad
\frac{1+\tr(\bar C)-2\bar{c}_1}{4},\quad
\frac{1+\tr(\bar C)-2\bar{c}_2}{4},\quad
\frac{1+\tr(\bar C)-2\bar{c}_3}{4}.
\end{equation}
Thus $\lambda(\bar\rho)=\frac{1}{4}\max[1-\tr(\bar C), 1+\tr(\bar C)-2\bar c_i]$. 

In the diagonal basis, the largest eigenvalue of $\bar \rho^{\rm diag}$ is $\lambda(\bar\rho^{\rm diag})=\frac{1}{4}\max[1-\tr(\bar C), 1+\tr(\bar C)-2\bar \mu_i]$.  From Eq.~\eqref{eq:mu1Ciimu3}, we have $\bar c_i \geq \bar \mu_3$ for any $i$. Then, $\lambda(\bar\rho^{\rm diag})\geq\lambda(\bar\rho) $ and $\mathscr{C}(\bar\rho^{\rm diag})\geq \mathscr{C}(\bar\rho)$.  Therefore, the diagonal basis maximizes the concurrence of angular-averaged states.

For leading-order $t\bar t$ production, the direction $\vec{n}$ in Fig.~\ref{fig:PlotCoordinates} is always an eigenvector of $C_{ij}$ (see App.~\ref{app:LzConservation} for details).  Whenever $\tr(C)<0$ or $C_{nn}$ is the smallest eigenvalue of $C_{ij}$,  the angular-averaged density matrix has exactly the same concurrence in the rotated beam basis, the helicity basis,  and the diagonal basis.  For $t\bar t$ production in QCD, this is the case for most channels (see Table~\ref{tab:qqggbasis}) and is the case in the near threshold region measured by ATLAS~\cite{ATLAS:2023fsd} and by CMS~\cite{CMS:2024pts}. Therefore, accidentally, the concurrence is usually not sensitive to basis choice as long as one axis is chosen as the normal direction of the scattering plane.

%%%%%%%%%%%%%%%%%%%%%%%%%%%%%%%%%%%%%%%%%%%%%%%%%%%%%%%%%%%
\subsection{Bell Inequality}
\label{sub:Bellinequalities}

We show next that the diagonal basis is also the one that maximizes Bell inequality violation for fictitious states.  Since the spin correlation matrix $\bar C$ of the angular-averaged sum $\bar \rho$ is symmetric, the eigenvalues of $(\bar C)^T\bar C$ are $\bar c_i^2$. In accordance with Eq.~(\ref{eq:Brho}), we have
\begin{equation}\label{eq:BelleigC}
\mathscr{B}(\bar \rho)= 2\max_{i\neq j}\sqrt{\bar c_i^2 + \bar c_j^2}.
\end{equation}
To show that the diagonal basis is the optimal basis for measuring the Bell inequality, \ie, $\mathscr{B}(\bar \rho^{\rm diag})\geq \mathscr{B}(\bar \rho)$,  we need to prove that for any $i\neq j$, there exist $k\neq \ell$ satisfying
\begin{equation}\label{eq:cicjlessthanmukmul}
\bar c_i^2+ \bar c_j^2 ~\leq~ 
\bar \mu_k^2+ \bar\mu_\ell^2,
\end{equation}
where $\bar\mu_i$ are the eigenvalues of $\bar C^{\rm diag}$, see Eqs.~\eqref{eq:Cdiagonal} and~\eqref{eq:mubardefinition}.

The relative signs of $\bar\mu_1$, $\bar\mu_2$, and $\bar \mu_3$ can be divided into the following three cases, and Eq.~\eqref{eq:cicjlessthanmukmul} can be proved case by case (see Ref.~\cite{Cheng:2023qmz} for details):
\begin{enumerate}[label=(\alph*)]
    \item $\bar\mu_1 \geq \bar\mu_2 \geq \bar\mu_3 \geq 0  \qquad \Longrightarrow\qquad \bar c_i^2+ \bar c_j^2 \leq \bar \mu_1^2 + \bar \mu_2^2 $,
    \item $0\geq \bar\mu_1 \geq \bar\mu_2 \geq \bar\mu_3 \qquad \Longrightarrow\qquad  \bar c_i^2+ \bar c_j^2 \leq \bar \mu_2^2 + \bar \mu_3^2 $,
    \item $\bar\mu_1 \geq 0 \geq \bar\mu_3 \qquad \Longrightarrow\qquad \bar c_i^2+ \bar c_j^2 \leq \bar \mu_1^2 + \bar \mu_3^2 $.
\end{enumerate}
Therefore, the fictitious state obtained using the diagonal basis gives the largest violation of Bell's inequality, $\mathscr B(\overline \rho^{\rm diag})\geq \mathscr B(\overline \rho)$.

%%%%%%%%%%%%%%%%%%%%%%%%%%%%%%%%%%%%%%%%%%%%%%%%%%%%%%%%%%%
\subsection{Alternative Bell Variable}

The variable $\mathscr{B}(\rho)$ is the theoretical upper limit of the Bell inequality violation as it is obtained by considering all possible values for the directions $\vec a_i$ and $\vec b_i$ to maximize Eq.~\eqref{eq:Bella1a2b1b2}.  In practice, for a simpler estimation of statistical uncertainties,\footnote{It is argued that the variable $\scrB(\rho)$ has asymmetric uncertainties which, if not accounted for properly, lead to a biased estimate~\cite{Severi:2021cnj}.} it is preferred to a priori fix the directions $\vec a_i$ and $\vec b_i$ with respect to the quantization axes as
\begin{equation}
\vec a_1 = \vec e_i, 
\quad 
\vec a_2 = \vec e_j, 
\quad 
\vec b_1=\frac{\vec e_i \pm \vec e_j}{\sqrt{2}}, 
\quad 
b_1=\frac{\vec e_i \mp \vec e_j}{\sqrt{2}},  
\quad 
(i\neq j).
\end{equation}
As a result, the observable for Bell inequality violation, Eq.~\eqref{eq:Bella1a2b1b2}, is
\begin{equation}\label{eq:BrhoUnbias}
\mathscr{B}'(\rho)=\sqrt{2}\max_{i\neq j} |C_{ii}\pm C_{jj}|.
\end{equation}
In the diagonal basis, the diagonal terms of the correlation matrix $\bar C_{ij}^{\rm diag}$ of the angular-averaged sum are simply the eigenvalues of $\bar C_{ij}^{\rm diag}$, \ie, $\bar C_{ii}^{\rm diag}=\bar\mu_i$.
In any other basis, as with Eqs.~\eqref{eq:trC1trC2} and \eqref{eq:mu1Ciimu3}, we have 
\begin{align}
    &\bar C_{11} + \bar C_{22} + \bar C_{33}=\bar \mu_1 + \bar \mu_2 + \bar \mu_3 = \tr(\bar C), \\ 
    \label{eq:Ciibound}
    &\bar \mu_1 \geq \bar C_{ii} \geq \bar \mu_3.
\end{align}
Here, Eq.~\eqref{eq:Ciibound} is satisfied because  $\bar C_{ii}$ are bounded by $\bar c_i$, the eigenvalues of $\bar C$, while $\bar c_i$ are bounded by $\bar \mu_i$.  Therefore the diagonal basis is also the one that maximizes the variable $\mathscr{B}'(\bar \rho)$.

%%%%%%%%%%%%
\subsection{Asymmetric Correlation Matrix}

If a $t\bar t$ event is constructed in an arbitrary Lorentz frame other than the $t\bar t$ center-of-mass frame, or if the production involves $CP$-violating interactions, the spin correlation matrix, $\bfC_\Omega$, for each quantum sub-state is generally not symmetric.  Still, the spin correlation matrix $\bfC_\Omega$ can be diagonalized with an event-dependent singular value decomposition:
\begin{equation}\label{eq:Csingular}
\bfC^{\rm diag}_\Omega=R_{\Omega}^{t}\bfC_\Omega \left(R^{\bar t}_{\Omega}\right)^T=
\begin{pmatrix}
\mu_{1,\Omega} &0&0\\
0&\mu_{2,\Omega} &0\\
0&0&\mu_{3,\Omega} \\
\end{pmatrix}.
\end{equation}
From $\bfC_\Omega$, the diagonal matrix $\bfC^{\rm diag}_\Omega$ can be obtained by using the singular vectors of $\bfC_\Omega$ to determine the two rotations, $R_{\Omega}^{t}$ and $R^{\bar t}_{\Omega}$, needed to rotate $\bfC_\Omega$ into $\bfC^{\rm diag}_\Omega$.  Let the three left (right) singular vectors of $\bfC_{\Omega}$ be 
$\vec{e}_{i,\Omega}^{\,t}$ ($\vec{e}_{i,\Omega}^{\, \bar t}$), then 
\begin{equation}
\vec{\, e}_{i,\Omega}^{\,t} = R_{ij,\Omega}^{t} \vec{e}_j,
\quad\quad\quad
\vec{\, e}_{i,\Omega}^{\,\bar t} = R_{ij,\Omega}^{\bar t} \vec{e}_j.
\end{equation}
Alternatively, within this Lorentz frame, one could measure the spins of $t$ and $\bar t$ with two different sets of quantization axes, $\vec{\, e}_{i,\Omega}^{\,t}$ and $\vec{\, e}_{i,\Omega}^{\,\bar t}$, to directly measure $\bfC^{\rm diag}_\Omega$.

Due to the transformation between $\bfC_\Omega$ and $\bfC^{\rm diag}_\Omega$ in Eq.~\eqref{eq:Csingular},  both spin correlation matrices have the same singular values.  Consequently, in any Lorentz frame the spin correlation matrix will have the same singular values and lead to the same matrix $\bfC^{\rm diag}_\Omega$ in the diagonal basis meaning that this generalized definition of the diagonal basis is independent of choice of Lorentz frame.  Our proof can be adjusted to include asymmetric correlation matrices (see App.~\ref{app:proof} for details) and thus we have shown that the diagonal basis is the optimal basis for general processes.

%%%%%%%%%%%%%%%%%%%%%%%%%%%%%%%%%%%%%%%%%%%%%%%%%%%%%%%%%%%
\section{Phenomenological Analyses for $t\bar t$ Events}
\label{sec:processes} 

In theory and in practice, it is perhaps most convenient to start the analysis for $t\bar t$ events in the helicity basis in the center-of-mass frame. The diagonal basis can be obtained from the helicity basis by a single rotation in the scattering plane.  This is because the vector $\vec n$ is always an eigenvector of the spin correlation matrix. This angle is shown visually in Fig.~\ref{fig:PlotCoordinates}(d) and relates the bases via
\begin{equation}
\vec{\,e}_1^{\,\rm diag}= \vec{r}\cos\xi+\vec{k}\sin\xi, \quad\quad\quad
\vec{\,e}_2^{\,\rm diag}= \vec{n},
\quad\quad\quad
\vec{\,e}_3^{\,\rm diag}= \vec{k}\cos\xi- \vec{r}\sin\xi.
\end{equation}
The rotation angle $\xi$ depends on the production process and on the phase space point of $t\bar t$. For simplicity, when we summarize the optimal (diagonal) basis for different SM processes in the following, all the correlation matrices are expressed in the helicity basis.

%%%%%%%%%%%%%%%%%%%%%%%%%%%%%%%%%%%%%%%%%%%%%%%%%%%%%%%%%%%
\subsection{Characteristics of QCD Sub-Processes}
\label{sec:qcd}

\begin{figure}
    \centering
    \includegraphics[width=0.7\linewidth]{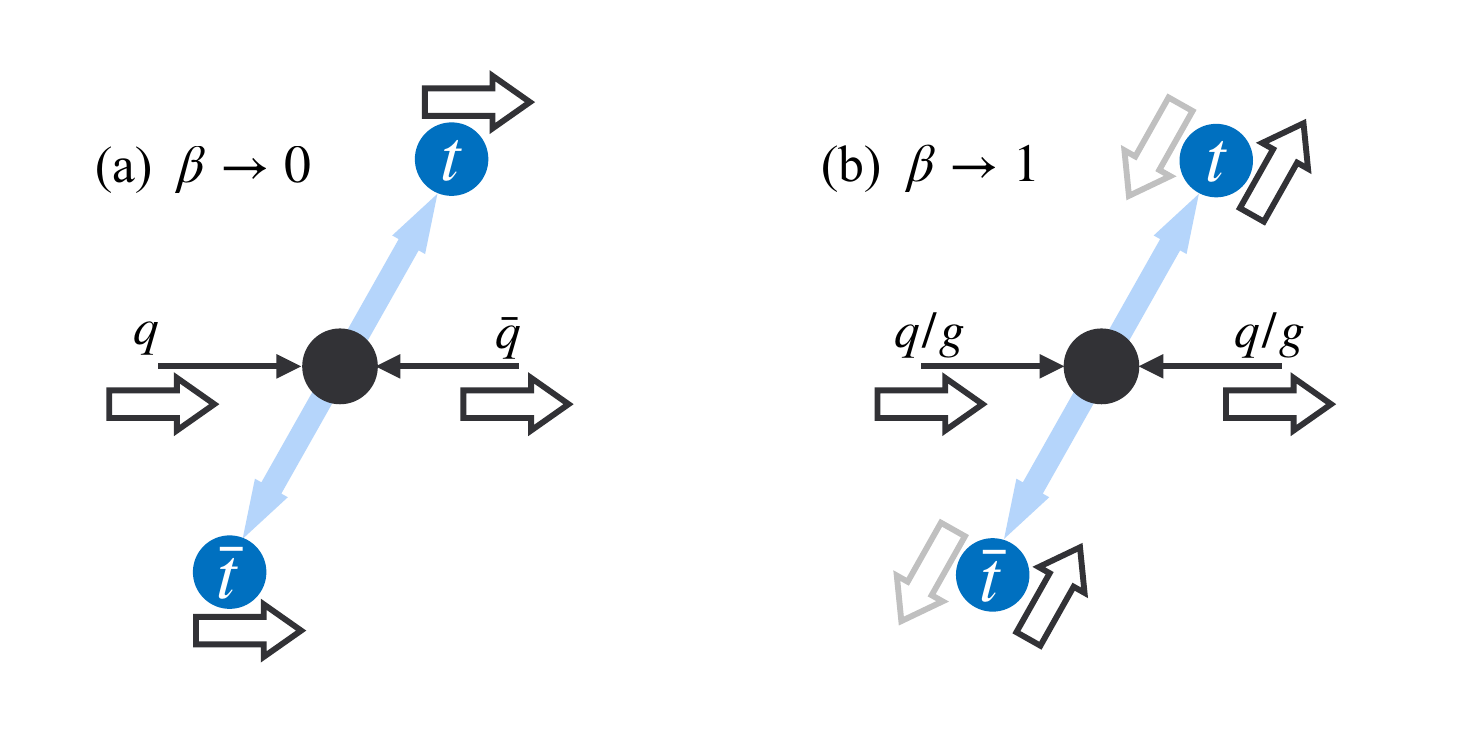}
    \caption{Spin configurations of $t\bar t$ produced from unlike-helicity initial states: (a) for $q\bar q\to t\bar t$ near threshold, with the cross section  proportional to $\beta$; and (b) for $q\bar q,\ g_L g_R\to t\bar t$ in the boosted region. Figure adapted from Ref.~\cite{Mahlon:2010gw}.}
    \label{fig:plotSpinForUnlikehelicity}
\end{figure}

\begin{figure}
    \centering
    \includegraphics[width=0.7\linewidth]{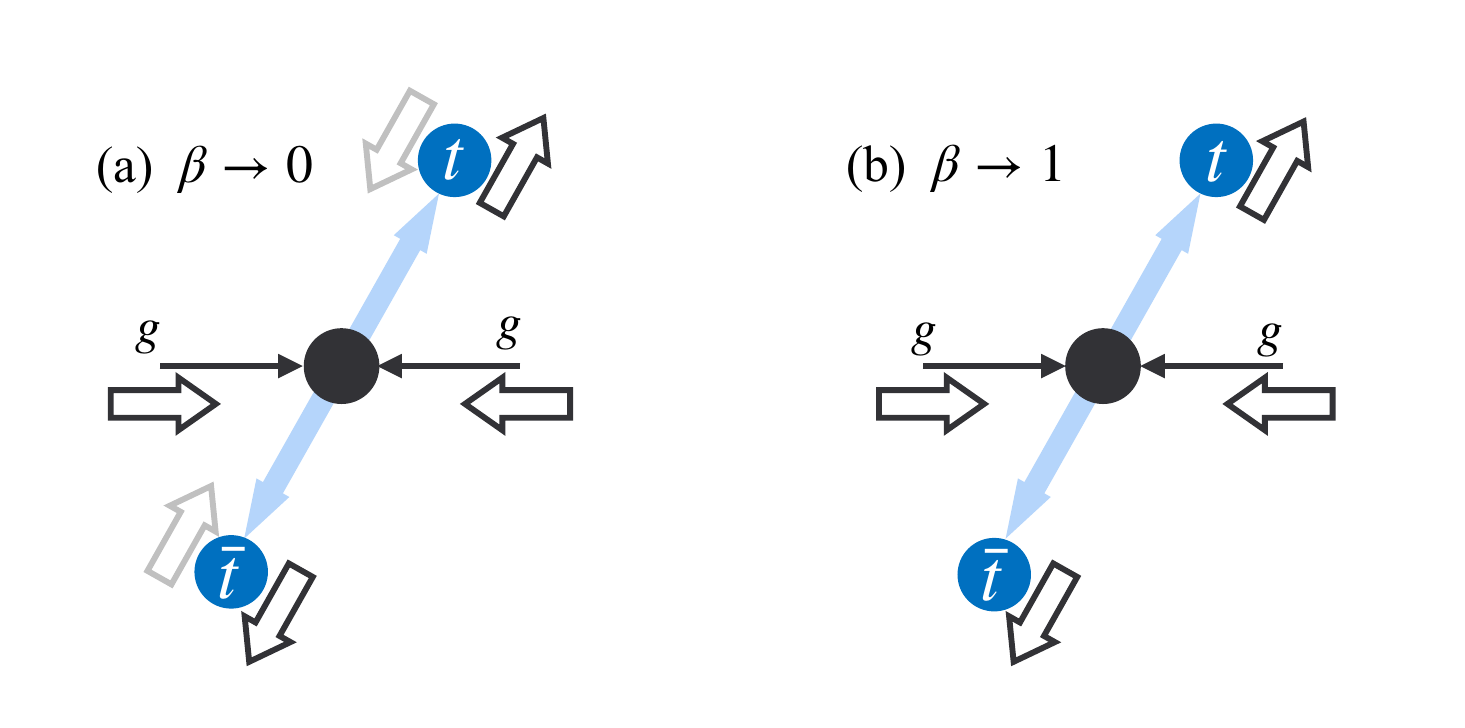}
    \caption{ Spin configurations of $t\bar t$ produced from like-helicity gluons near and above threshold. The cross section is proportional to $\beta$. Figure adapted from Ref.~\cite{Mahlon:2010gw}.}
    \label{fig:plotSpinForLikehelicity}
\end{figure}

We first consider the parton-level processes in QCD of $q\bar q\to t\bar t$ and $gg\to t\bar t$. We express the $2\to 2$ unpolarized production cross section at the partonic level as
\begin{equation}
\operatorname{d} \sigma(t\bar t) = \frac{\beta} {64\pi^2 \hat{s}}\
    \overline\sum |\mathcal M|^2\ \operatorname{d}\Omega,
    \quad\quad\quad
\operatorname{d}\Omega= 
    \operatorname{d}\phi\ 
    \operatorname{d}\cos\theta. 
    \nonumber
\end{equation}
With respect to the $t\bar t$ spin states labeled by $\alpha,\beta$, the scattering matrix element and the spin density matrix in Eq.~\eqref{eq:rhoBiCij} are related in the $4\times 4$ matrix form as
\begin{equation}
 {\sum}|\mathcal{M}|^2 =\tr(\mathcal{M}_{\alpha\bar\alpha} \mathcal{M}^*_{\beta\bar\beta})\equiv  \tr(\mathcal{R}_{\alpha\beta,\bar\alpha\bar\beta}),
 \quad\quad\quad
 \rho_{\alpha\beta,\bar\alpha\bar\beta} = \frac{ \cal{R}_{\alpha\beta,\bar\alpha\bar\beta}} { \tr(\cal{R}) } .
 \nonumber 
\end{equation}
Further details can be found in App.~\ref{app:basistransform}.  As seen in Eq.~\eqref{eq:rhoBiCij}, $B^\pm_i$ terms do not contribute for an unpolarized $t \bar t$ system, and the density matrix is thus obtained by summing over the correlation matrix elements $C_{ij}$.  The squared matrix elements, spin correlation matrices $C_{ij}$, and the rotation angle $\xi$ of the optimization for each process are listed in Table~\ref{tab:qqggbasis}, where the velocity factor in the center-of-mass frame of $t\bar t$ is $\beta=(1 -4m_t^2/\hat{s})^{1/2}$ and the boost factor is $\gamma=\sqrt{\hat{s}}/(2m_t)$, where $\sqrt{\hat{s}}$ is the center-of-mass energy of $t\bar t$.  The QCD dynamical pre-factors from the tree-level matrix element calculations are 
\begin{equation}\label{eq:kappaQCD}
\kappa_q=g_s^2 \frac{(N^2-1)}{N^2},
\quad\quad\quad
\kappa_g=\frac{2g_s^2}{(\beta^2\cth^2-1)^2}\frac{N^2(\beta^2\cth^2+1)-2}{N(N^2-1)},
\end{equation}
where $N=3$ is the number of colors, and $\cth = \cos\theta$ is the scattering angle of the top quark in the $t\bar t$ center-of-mass frame; see Fig.~\ref{fig:PlotCoordinates}.  
The spin configurations of $t\bar t$ listed in Table~\ref{tab:qqggbasis} are illustrated in Fig.~\ref{fig:plotSpinForUnlikehelicity} and Fig.~\ref{fig:plotSpinForLikehelicity}, where the $g_L g_R \to t\bar t$ process is $\beta^3$-suppressed near threshold and therefore omitted in Fig.~\ref{fig:plotSpinForUnlikehelicity}(a).  The $t\bar t$ spins in Fig.~\ref{fig:plotSpinForLikehelicity}(a) have no preferred orientation.

\begin{table}
  \centering
  \begin{tabular}{|c|c|c|c|}
  \hline
        QCD production & $\overline\sum |\mathcal M|^2$ & spin correlation matrix $C_{ij}$& $\xi$  \\
    \hline
        $q\bar q\to t\bar t$ & $\kappa_q\left(2-\beta^2\sth^2\right)$ & \multirow{2}{*}{$\Cijqqtt$} & $\tan\xi=\frac{1}{\gamma}\tan\theta$ \rule[-18pt]{0pt}{40pt} \\
    \cline{1-2} \cline{4-4}
        $g_Lg_R\to t\bar t$ &  $\kappa_g \beta^2\sth^2(2-\beta^2\sth^2)$ &  & $\tan\xi=\frac{1}{\gamma}\tan\theta$ \rule[-18pt]{0pt}{40pt} \\
    \hline
        $g_Lg_L/g_Rg_R\to t\bar t$ &  $\kappa_g (1-\beta^4)$ & $\CijggttSzero$ & $\xi=0$ \\
    \hline
    \end{tabular}
    \caption{The squared matrix element, spin correlation matrix $C_\Omega$, and rotation angle to the diagonal basis for $t\bar t$ production for the initial states of $q\bar q$, unlike-helicity gluons ($g_L g_R$), and like-helicity gluons ($g_Lg_L/g_Rg_R$).  The spin correlation matrix is expressed in the helicity basis ($\vec{r}, \vec{n}, \vec{k}$).}
    \label{tab:qqggbasis}
\end{table}

%%%%%%%%%%%%%%%%%%%%%%%%%%%%%%%%%%%
\subsubsection{$q\bar{q}\to t\bar{t}$ Process} 

For $q\bar q\to t\bar t$, the differential value of the spin correlation matrix in the diagonal basis $(\vec{e}_1^{\rm\, diag},\vec{e}_2^{\rm\, diag}, \vec{e}_3^{\rm\, diag})$ is
\begin{equation}
C^{\rm diag}=
\begin{pmatrix}
    \frac{\beta^2\sth^2}{2-\beta^2\sth^2} &0&0 \\
    0&-\frac{\beta^2\sth^2}{2-\beta^2\sth^2} &0 \\
    0&0&1 \\
\end{pmatrix}.
\end{equation}
The spins of the top and anti-top along the third direction $\vec{e}_3^{\rm \, diag}$ are 100\% correlated.  This reproduces the well-known result in Ref.~\cite{Mahlon:1995zn}.  Such constructed $t\bar t$ quantum sub-states produced from $q \bar q\to t\bar t$ at any fixed phase space point are entangled.  In particular, in the high-$p_T$ region where $\beta\approx1$ and $s_\theta\approx 1$, the spins of the top and the anti-top are maximally correlated along all three directions, resulting in a maximally entangled state.  The maximal violation of the Bell inequality is given by the largest two eigenvalues of the spin correlation matrix as seen in Eq.~\eqref{eq:BelleigC}.

For $q\bar q\to t\bar t$, the angle between the diagonal basis and the helicity basis (see Fig.~\ref{fig:PlotCoordinates}) is $\xi=(\tan\theta)/\gamma$.  Near threshold $\gamma \to 1$ and the rotation angle $\xi$ is approximately the same as the scattering angle $\theta$, which means that the diagonal basis approaches the rotated beam basis.  In the highly-boosted region, the rotation angle $\xi$ becomes zero and the diagonal basis coincides with the helicity basis.  To show how the optimal (diagonal) basis maximizes the Bell inequality violation, we average the quantum sub-states over all of phase space, and compare the Bell inequality violation of the states averaged in different bases. 

\begin{figure}
  \centering
  \includegraphics[width=.45\linewidth]{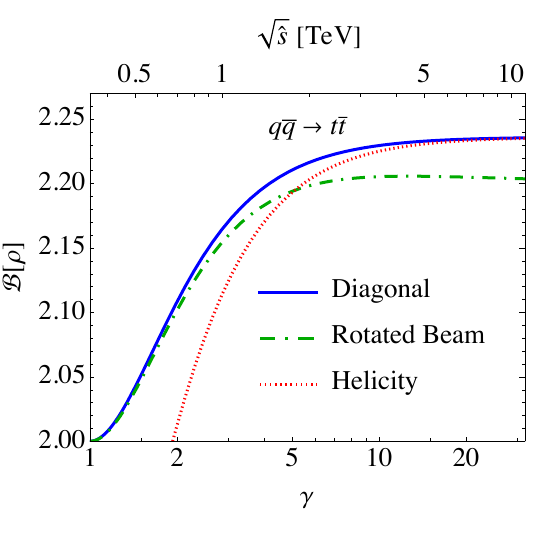}  
  \caption{Bell inequality violation of angular-averaged states as a function of center-of-mass energy.  Bell inequality violation occurs when $\mathscr{B}[\rho]>2$. }
  \label{fig:comparebasisqq}
\end{figure}

We present our numerical results in Fig.~\ref{fig:comparebasisqq} in the rotated beam basis, the helicity basis, and the diagonal basis. We recall that Bell inequality violation is inferred when $\mathscr B >2$. We see that the optimal (diagonal) basis usually leads to a considerable increase of Bell inequality violation, reaching ${\mathscr B} \approx 2.24$ far above threshold. In the boosted region or near threshold, the commonly-used helicity basis or rotated beam basis, respectively, are already nearly optimal.  We also evaluated the density matrix in the fixed beam basis with the azimuthal angle averaged, but the effects for Bell inequality violation are hardly visible in this basis, as explained in detail in App.~\ref{app:LzConservation}.   

%%%%%%%%%%%%%%
%%%%%%%%%%%%%%

%%%%%%%%%%%%%%%%%%%%%%%%%%%%%%%%%%
\subsubsection{$gg \to t\bar t$ Process} 

The process $gg\to t\bar t$ can be divided into two sub-processes:  unlike-helicity gluon scattering ($g_L g_R$, $P$-wave, with angular momentum $|S_z|=2$) and like-helicity gluon scattering ($g_L g_L, g_R g_R$, $S$-wave, with angular momentum $|S_z|=0$), as shown in Table~\ref{tab:qqggbasis}, Fig.~\ref{fig:plotSpinForUnlikehelicity}, and Fig.~\ref{fig:plotSpinForLikehelicity}. 

The top pair produced from unlike-helicity gluons has exactly the same spin correlation as in the $q\bar q \to t \bar t$ process.  Consequently, the $t$ and $\bar t$ usually have like-spins, \eg, with $\vec k$ as spin quantization axis, the $t\bar t$ spin state is $(\ket{\uparrow\uparrow}+\ket{\downarrow\downarrow})/\sqrt{2}$, in the high-$p_T$ region. The $t\bar t$ produced from like-helicity gluons, on the other hand, usually have unlike-spins. These like-helicity gluons produce the $t$ and $\bar t$ in a spin singlet state $(\ket{\uparrow\downarrow}-\ket{\downarrow\uparrow})/\sqrt{2}$.  This is a maximally-entangled configuration, especially near threshold.

While the top pair produced from either like-helicity or unlike-helicity gluons are entangled as spin singlet or triplet configuration, the complete process of $gg\to t\bar t$ with unpolarized gluons does not always result in an entangled $t\bar t$ system.  In general, the top pair is produced in a mixed state with a spin correlation of
\begin{equation}
C_{gg\to t\bar t} = \frac{|\mathcal{M}_{\text{unlike}}|^2 C_{\text{unlike}} + |\mathcal{M}_{\text{like}}|^2 C_{\text{like}}}{|\mathcal{M}_{\text{unlike}}|^2+|\mathcal{M}_{\text{like}}|^2} \ .
\end{equation}
The squared amplitudes $|\mathcal{M}_{\text{(un)like}}|$ and spin correlation matrices $C_{\text{(un)like}}$ for like-helicity and unlike-helicity gluon scattering are listed in Table~\ref{tab:qqggbasis}.  Comparing the squared amplitudes, we find that like-helicity gluon scattering dominates when $\beta \ll 1$ near the production threshold, while unlike-helicity gluon scattering dominates when $\gamma s_\theta\gg1$, far above threshold.  Consequently, top pairs produced from gluon scattering are entangled in these two separated kinematic regions: near threshold and high-$p_T$ region.

\begin{figure}
  \centering
  \includegraphics[width=.45\linewidth]{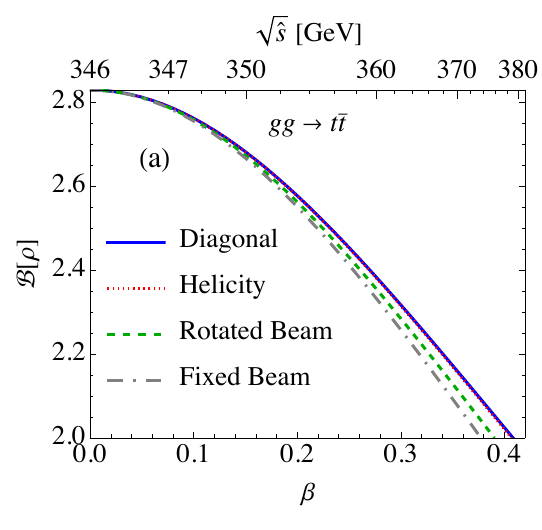}  
  \includegraphics[width=.45\linewidth]{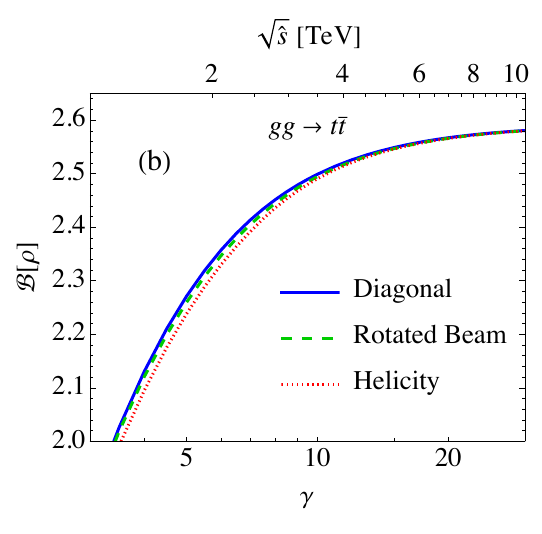}  
  \caption{Bell inequality violation of $gg\to t\bar t$. (a) Near threshold; the density matrix is angular-averaged. (b) High-$p_T$ region with a selection cut $|\cos\theta|<0.5$ imposed.}
  \label{fig:comparebasisgg}
\end{figure}

We present numerical results for Bell inequalities at the partonic level for $gg\to t \bar t$ in Fig.~\ref{fig:comparebasisgg} for various basis choices.  Near threshold, the top pair produced from like-helicity gluon scattering approximately forms a spin singlet. The correlation matrix is proportional to the identity matrix at leading order, $C_{ij}={\rm diag}(-1,-1,-1)+\mathcal{O}(\beta^2)$, which leads to maximal Bell inequality violation reaching ${\cal B} = 2\sqrt{2}$ when $\beta \to 0$, as seen in Fig.~\ref{fig:comparebasisgg}(a). The entanglement of $t\bar t$ produced from $gg$ is not sensitive to the basis choice in this case.   We note that $gg\to t\bar t$ is significantly more sensitive than $q\bar q \to t \bar t$ in measuring entanglement and Bell inequality violation near threshold. 

In the high-$p_T$ region where $\gamma\sth\gg 1$ ($\gamma\gg1$ and $c_\theta\approx 0$), the top pair produced from unlike-helicity gluon scattering forms a spin triplet, and the spin correlation matrix is $C_{ij}={\rm diag}(1,-1,1)+\mathcal{O}(\cth^2)+\mathcal{O}(\gamma^{-2})$.  The spin correlation matrix is approximately invariant under any rotation in the scattering plane.  After fixing one quantization axis $\vec e_2=\vec{n}$, the entanglement is not sensitive to the choice of the other two axes. As a result, the fictitious states  obtained in both the helicity basis and the rotated beam basis both yield nearly the optimal signal of Bell inequality violation, reaching ${\cal B} \approx 2.6$ at large $\gamma$, as seen in  Fig.~\ref{fig:comparebasisgg}(b).  Similar to $q\bar q\to t\bar t$, in the high-$p_T$ region, no Bell inequality violation can be seen if the density matrix is averaged over the azimuthal angle in the fixed beam basis.  

It is important to note that the like-helicity and unlike-helicity gluons contribute oppositely to the spin correlation of the top pair, \ie, top pairs produced from like-helicity gluons ($g_L g_L, g_Rg_R$) tend to have like-spins while top pairs produced from unlike-helicity gluons ($g_Lg_R$) tend to have unlike-spins; see Figs.~\ref{fig:plotSpinForUnlikehelicity} and~\ref{fig:plotSpinForLikehelicity} for illustration.  As a result, in the phase space region where the contribution like-helicity and unlike-helicity gluons are comparable, the spin correlation of $t\bar t$ is largely canceled and there is no entanglement in $t\bar t$ system.
Similarly, the spin correlation of $t\bar t$ produced from $q\bar q$ and like-helicity gluons also cancel with each other.

%%%%%%%%%%%%%%%%%%%%%%%%%%%%%%%%%%%%%%%%%%%%%%%
\subsection{Optimization Procedure at the LHC}

\begin{figure}[tb]
  \centering
  \includegraphics[height=0.42\linewidth]{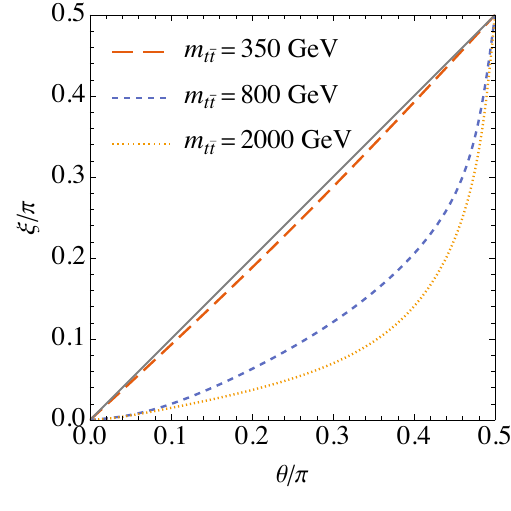}
  \caption{The rotation angle $\xi$ that obtains the optimal (diagonal) basis as a function of scattering angle $\theta$ for different values of $m_{t\bar t}$.}
  \label{fig:LHCoptimal}
\end{figure}

At hadron colliders, in reality, the $t\bar t$ quantum state is a mixed state due to contributions from both $q\bar q \to t\bar t$ and $gg\to t\bar t$
\begin{equation}
\rho^{t\bar t} = \omega^{q\bar q} \rho^{q\bar q\to t\bar t} + \omega^{gg} \rho^{gg\to t\bar t}.
\end{equation}
At a given center-of-mass energy $\sqrt{\hat s}$, the weights $\omega^{q\bar q}$ and $\omega^{gg}$ are determined by the parton luminosities and production rates
\begin{equation}
\omega^{I}=\frac{L_{I} |\mathcal{M}_{I\to t\bar t}|^2}{L_{q\bar q}|\mathcal{M}_{q\bar q\to t\bar t}|^2 +L_{gg}|\mathcal{M}_{gg\to t\bar t}|^2},
\quad\quad\quad
I=q\bar q,\  gg.
\end{equation}
Taking the spin correlation matrices from Table~\ref{tab:qqggbasis} into account, we obtain the diagonal basis for realistic hadronic processes
\newcommand{\factorq}{(L_{qq}\kappa_q+L_{gg}\kappa_g\beta^2\sth^2)}
\begin{equation}\label{eq:tan2xipp}
\tan(2\xi)
=
\frac{\factorq s_{2\theta}\sqrt{1-\beta^2}}{\factorq (c_{2\theta}+\beta^2\sth^2) + L_{gg}\kappa_g \beta^2(1-\beta^2)},
\end{equation}
where $\kappa_{q}$ and $\kappa_g$ are the QCD factors defined in Eq.~\eqref{eq:kappaQCD}.   This result generalizes the basis of the $gg\to t\bar t$ process obtained in Refs.~\cite{Uwer:2004vp,Mahlon:2010gw}.  In Fig.~\ref{fig:LHCoptimal}, we show the relation between the angles $\xi$ and $\theta$ at the $13~\tev$ LHC for various values of the $t\bar t$ invariant mass.  Near threshold at low $m_{tt}$ and $\beta \to 0$, $\tan(2\xi) \approx \tan(2\theta)$; thus the rotated beam basis aligns with the optimal (diagonal) basis. Far above threshold when $\beta\to 1$, $\tan(2\xi) \to 0$ and the helicity basis aligns with the optimal (diagonal) basis. 

    We obtain the optimal basis given by Eq.~\eqref{eq:tan2xipp} in the leading order kinematics, which may receive event-dependent corrections at higher orders. The complete full order treatments are beyond the scope of this work. The NLO corrections of the correlation matrix as a function of $m_{t\bar t}$ and $\theta$~\cite{Frixione:2007nw,Nason:2004rx,Frixione:2007vw,Alioli:2010xd} have been applied to the analyses of ATLAS and CMS, where the spin correlation matrix in the optimal basis choice from Eq.~\eqref{eq:tan2xipp} will only be approximately diagonal. One may consider vetoing the additional jet activity in the events to recover the simple leading order kinematics.

%%%%%%%%%%%%%%
%%%%%%%%%%%%%%
%

\begin{figure}[tb]
    \centering
    \includegraphics[height=0.42\linewidth]{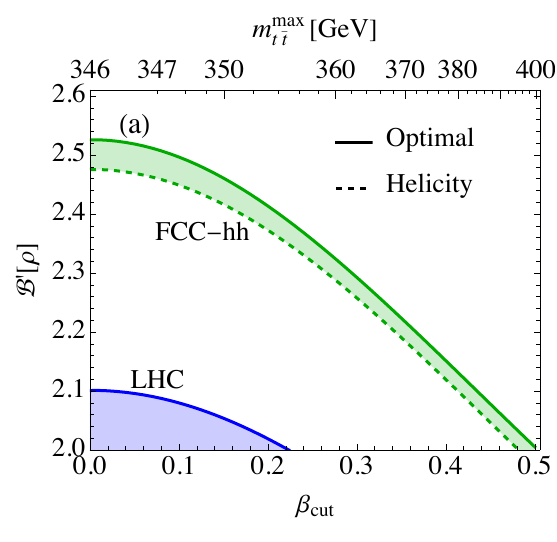} 
    \includegraphics[height=0.42\linewidth]{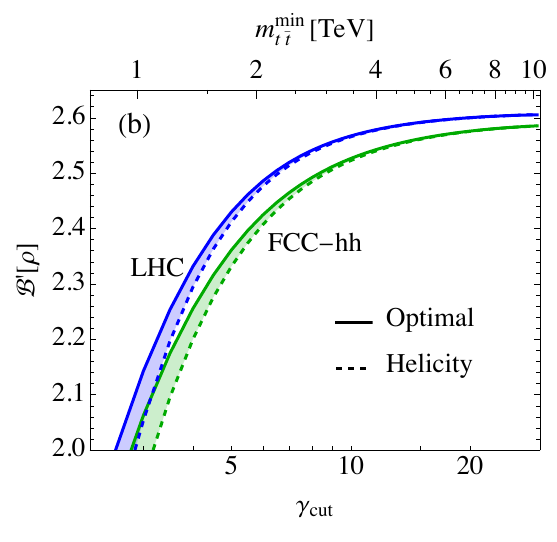} 
    \caption{Bell inequality violation $\mathcal{B}'$ of $pp\to t\bar t$.
    (a) Results near threshold versus upper cut of $m_{t\bar t}$; the density matrix is fully angular-averaged.  Note that $\mathcal{B}'[\bar\rho^{\rm hel}]<2$ for the state averaged in helicity basis at LHC.  (b) Result in high $p_T$ region versus lower cut of $m_{t\bar t}$; an angular selection cut $|\cos\theta|<0.5$ is chosen.}
    \label{fig:comparebasisLHC}
\end{figure}

At the LHC energies, $t\bar t$ production is dominated by gluon fusion.  Both the near-threshold and the high-$p_T$ regions contribute uniquely to the quantum information properties.  Near threshold, top pairs produced from gluon fusion are entangled in a spin singlet configuration. However, the contribution of $q\bar q$ can have a sizable effect on the Bell inequality violation of $t\bar t$.  As illustrated in Figs.~\ref{fig:plotSpinForUnlikehelicity} and~\ref{fig:plotSpinForLikehelicity}, the spin correlation of a $t\bar t$ pair produced from unlike-helicity gluons and $q\bar q$ have opposite signs and cancel with each other.  This cancellation greatly reduces the signal of entanglement and makes measuring Bell inequality violation near threshold challenging at the LHC.  We show the Bell variable both near the threshold versus a upper cut on the speed $\beta$, in Fig.~\ref{fig:comparebasisLHC}(a), and above threshold versus a lower cut on $\gamma$, in Fig.~\ref{fig:comparebasisLHC}(b), which are equivalently to an upper and a lower bound on the invariant mass $m_{t\bar{t}}$ (labeled by the upper horizontal axes), respectively.  We calculate the parton luminosity with \texttt{CT18}~\cite{Hou:2019efy} and demonstrate the results in Fig.~\ref{fig:comparebasisLHC}(a), where the Bell inequality violation decreases to about 2.11 at the energy of the LHC even though the gluon fusion still dominates $t\bar t$ production ($L_{q\bar q}/L_{gg}\approx 0.11$).  For illustration, we also calculate the expectation at a future 100 TeV hadron collider (FCC-hh), where we achieve a result of about 2.55 near threshold because the $q\bar q$ contribution is further suppressed relative to the LHC ($L_{q\bar q}/L_{gg}\approx 0.04$).  It is also important to note that the optimal (diagonal) basis approaches the rotated beam basis near threshold.  In particular, when using the Bell variable $\mathcal{B}'$ defined in Eq.~\eqref{eq:BrhoUnbias}, the violation of Bell inequalities near threshold is only measurable in the beam basis but not in the helicity basis.  This explains the simulation results in Ref.~\cite{Aguilar-Saavedra:2022uye}.

In the high-$p_T$ region, top pairs produced at hadron colliders are entangled as the $t\bar t$ are in a spin triplet configuration.  Both quark initial states and unlike-helicity gluon initial state contribute to the spin triplet, while the like-helicity gluon scattering contaminates the spin correlation as summarized at the end of Sec.~\ref{sec:qcd}.  Therefore, more $q\bar q\to t\bar t$ contribution makes the Bell inequality violation larger, but generally the result is not highly sensitive to $q\bar q$ contribution as the contribution from like-helicity gluon scattering is small at high-$p_T$ region; see Fig.~\ref{fig:comparebasisLHC}(b).

%%%%%%%%%%%%%%%%%%%%%%%
\subsection{Optimization Procedure at Lepton Colliders}

At a lepton collider such as an $e^+ e^-$ collider,  top pair production not far above the threshold proceeds through the $s$-channel process $e^+ e^-\to t\bar t$ via the $\gamma/Z$ exchange.  This process generally produces the same spin correlations as the process $q\bar q \to t\bar t$. 
 The diagonal basis was first explored in $e^+ e^-\to t\bar t$ in Ref.~\cite{Parke:1996pr}, where it was called the ``off-diagonal basis.'' 

The difference between electroweak processes and QCD processes is that the $Z$ boson couplings to fermions are chiral so the electroweak process $e^+ e^- \to t\bar{t}$ always results in a certain degree of polarization even if the incoming beams are unpolarized.

Including both the vector and axial current interactions, a general $s$-channel amplitude can be parametrized as 
\begin{equation}
\mathcal{M}_{t\bar{t}} =  \frac{e^2g_{\mu\nu}}{s} J_{\rm in, \pm 1}^\mu \left( f_V J_{{\rm out},V}^\nu + f_A J_{{\rm out},A}^\nu  \right).
\end{equation}
Here, $J^\mu_{\rm in,\pm1}=\frac{\sqrt{\hat{s}}}{2}(0,1,\pm i,0)$ is the current of initial states, $J^\mu_{{\rm out},V}=\bar t\gamma^\mu t$ and $J^\mu_{{\rm out},A}=\bar t\gamma^\mu\gamma^5 t$ are vector and axial currents of the final states, respectively. $f_{V/A}$ parametrizes the interaction strength of the vector/axial-vector current. More specifically, the amplitudes of $e^+ e^- \to t\bar{t}$ are written as 
\begin{align}
    \mathcal{M}_{e^-_L e^+_R \to t\Bar{t}} &=  \frac{e^2g_{\mu\nu}}{s} J_{\rm in, -1}^\mu \left( f^L_V J_{{\rm out},V}^\nu + f^L_A J_{{\rm out},A}^\nu  \right),\\
    \mathcal{M}_{e^-_R e^+_L \to t\Bar{t}} &=  \frac{e^2g_{\mu\nu}}{s} J_{\rm in, 1}^\mu \left( f^R_V J_{{\rm out},V}^\nu + f^R_A J_{{\rm out},A}^\nu  \right),
\end{align}
and the coupling strengths are
\begin{align}
    %g_{LL}&= Q_e Q_t + (I^3_e-Q_e s_W^2) (I^3_t-Q_t s_W^2) \frac{1}{c_W^2 s_W^2} \frac{\hat{s}}{\hat{s}-m_Z^2} \\
    %g_{LR}&= Q_e Q_t + (I^3_e-Q_e s_W^2) (-Q_t s_W^2) \frac{1}{c_W^2 s_W^2} \frac{\hat{s}}{\hat{s}-m_Z^2}\\
    f_{V}^L&= Q_e Q_t +  \frac{(I^3_e-Q_e s_W^2) (I_t^3-2Q_t s_W^2)}{2c_W^2 s_W^2} \frac{\hat{s}}{\hat{s}-m_Z^2},\\
    f_{A}^L&= -  \frac{(I^3_e-Q_e s_W^2) I_t^3}{2c_W^2 s_W^2} \frac{\hat{s}}{\hat{s}-m_Z^2},\\
    %g_{RL}&= Q_e Q_t + (-Q_e s_W^2) (I^3_t-Q_t s_W^2) \frac{1}{c_W^2 s_W^2} \frac{\hat{s}}{\hat{s}-m_Z^2} \\
    %g_{RR}&= Q_e Q_t + (-Q_e s_W^2) (-Q_t s_W^2) \frac{1}{c_W^2 s_W^2} \frac{\hat{s}}{\hat{s}-m_Z^2}
    f_{V}^R&= Q_e Q_t  - \frac{Q_e (I_t^3-2Q_t s_W^2)}{2c_W^2} \frac{\hat{s}}{\hat{s}-m_Z^2},\\
    f_{A}^R&=  \frac{Q_e I_t^3}{2c_W^2} \frac{\hat{s}}{\hat{s}-m_Z^2},
\end{align}
where $Q_e=-1$, $Q_t=2/3$, $I_e^3=-1/2$, and $I_t^3=1/2$. The results above can be directly generalized to the pair production of any fermion. For instance, for the $\tau^-\tau^+$ system, we simply  replace  $Q_t$ with $Q_\tau=-1$ and $I_t^3$ with $I^3_\tau=-1/2$.

Since the tops are produced with some polarization, the $B_i^\pm$'s in Eq.~\eqref{eq:rhoBiCij} are non-zero, and the concurrence cannot be simply expressed by the eigenvalue of $C_{ij}$ as in section~\ref{sub:concurrence}.  It is thus important to measure all the parameters $B_i^\pm$ and $C_{ij}$, to achieve quantum tomography for the $t\bar t$ bipartite state.  On the other hand, Bell's inequality only depends on the correlation matrix $C_{ij}$ and our discussion in  Sec.~\ref{sub:Bellinequalities} is still valid.

\begin{figure}[t]
    \centering   \includegraphics[height=0.42\linewidth]{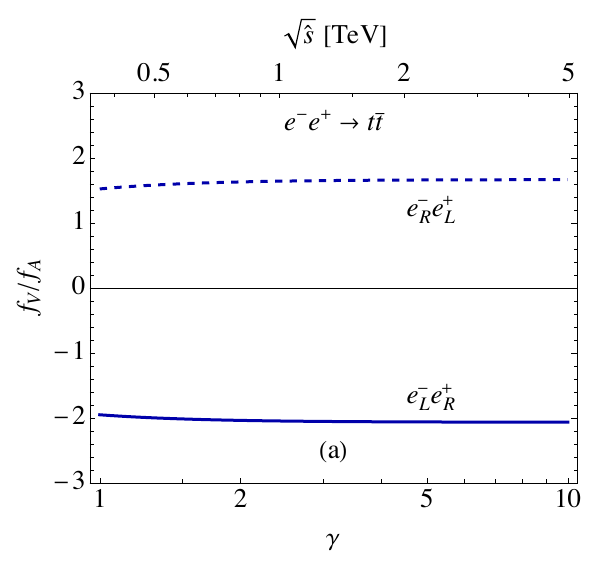}    \includegraphics[height=0.42\linewidth]{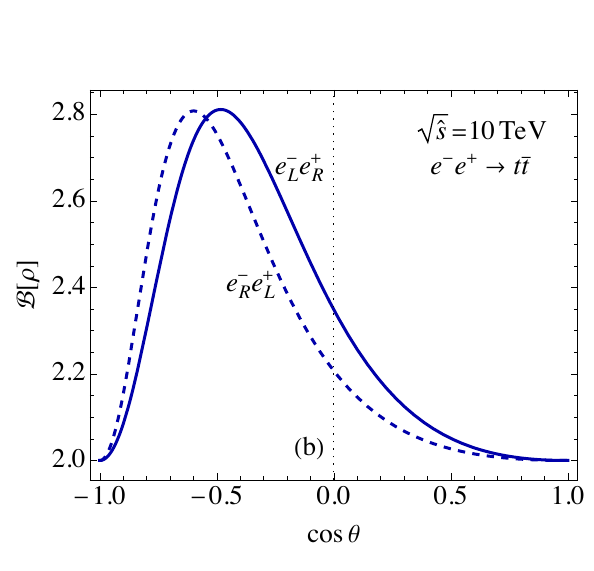}
    \caption{ (a) The ratio between vector and axial coupling strength as a function of energy. (b) The Bell variable $\mathcal{B}[\rho]$ as a function of $\cos\theta$ at when $\sqrt{s}=10$ TeV. }
    \label{fig:fvfa}
\end{figure}

For the $e^-_Re^+_L\to t\bar t$ and $e^-_Le^+_R\to t\bar t$ processes, respectively, the spin correlation matrix of $t\bar t$ is given in terms of the chiral couplings and the scattering angle $\theta$ in the center-of-mass frame in the helicity basis as
\begin{align}
    &C_{ij}=\frac{1}{\Cijeedenomiator{f_V}{f_A}{\pm}} \times \\
    &\Cijeett{f_V}{f_A}{\pm}.    \nonumber
\end{align}
The eigenvalues $\mu_i$ of the spin correlation matrix are
\begin{equation}\label{eq:mu123ee}
\mu_1=-\mu_2=\frac{\beta^2\sth^2(f_V^2-f_A^2)}{f_V^2(2-\beta^2\sth^2)\pm 4\beta f_V f_A \cth + \beta^2 f_A^2 (1+\cth^2)},
\quad\quad\quad
\mu_3=1,
\end{equation}
and the maximal value of $|\mu_1|$ is  reached at an angle $\cos\theta = \mp \beta f_A/ f_V$.  The spin correlation matrices of $t\bar t$ produced from $e_R^- e_L^+$ and $e_L^- e_R^+$ are related by $f^R_V\leftrightarrow f^L_V$ and $f^R_A\leftrightarrow -f^L_A$.  These two processes with different electron polarizations lead to similar $t\bar t$ angular distributions and spin correlations~\cite{Parke:1996pr} because numerically $f^R_V/f^R_A \sim -f^L_V/f^L_A$.  Additionally, these ratios have a mild dependence on the collision energy, as shown in Fig.~\ref{fig:fvfa}(a).  As a result, the quantum entanglement of a top pair produced at $e^+ e^-$ colliders is not very sensitive to electron beam polarization. We show this by plotting the Bell variable $\mathcal{B}[\rho]$ for quantum sub-states produced at different scattering angles in Fig.~\ref{fig:fvfa}(b).  On the other hand, the ratio between vector and axial current coupling strength can be changed by new physics (such as models with new vector bosons), and the entanglement of top pairs at $e^+e^-$ colliders is sensitive to such new physics. 

%%%%%%%%%%%%%%%%%%%%%%%%
%%%%%%%%%%%%%%
\begin{figure}[t]
  \centering 
\includegraphics[height=0.48\linewidth]{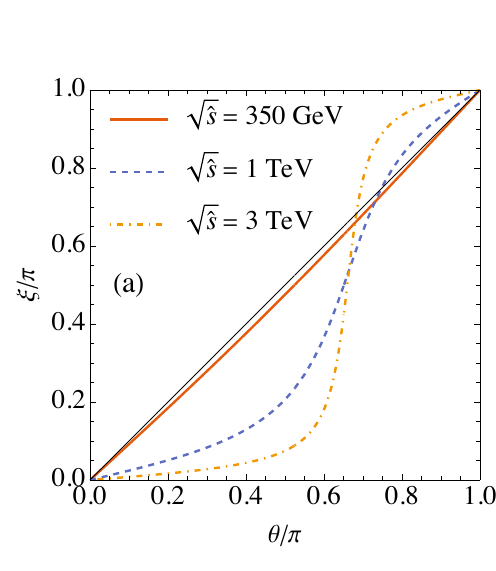}  
  \quad\quad
\includegraphics[height=0.47\linewidth]{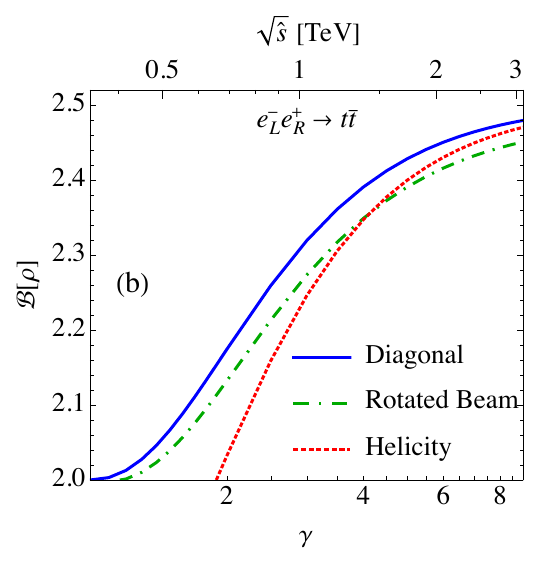}  
  \caption{(a) 
  The rotation angle $\xi$ that obtains the optimal (diagonal) basis as a function of scattering angle $\theta$ for different values of $\sqrt{\hat s}$.  (b)  Bell inequality violation of fictitious states averaged in different bases.  A selection cut of $\cos\theta<0$ is used to choose the mostly entangled phase space. }
  \label{fig:comparebasisee}
\end{figure}
%%%%%%%%%%%%%%

\begin{figure}
    \centering  
    \includegraphics[width=.45\linewidth]{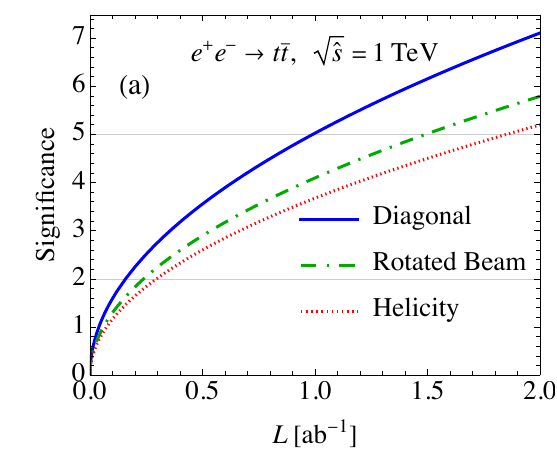} \quad\quad    \includegraphics[width=.45\linewidth]{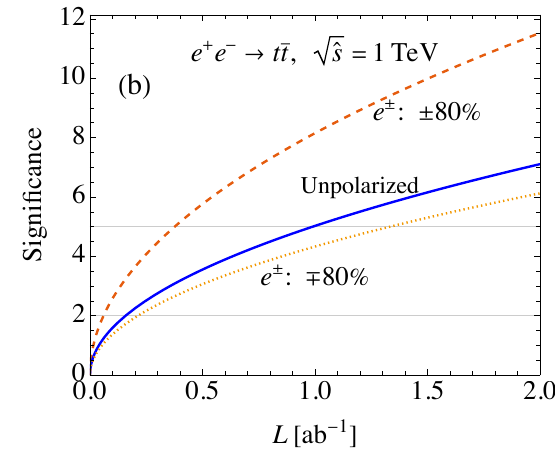}
    \caption{Significance of measuring $\mathscr{B}(\rho)>2$ at 1 TeV $e^+ e^-$ colliders assuming 100\% efficiency. (a) Different basis choices for unpolarized scattering. (b) Optimal (diagonal) basis for different beam polarizations 
    with the dashed curve for  $(e^+,e^-=+80\%, -80\%)$ and dotted curve for $(e^+,e^-=-80\%, +80\%)$, where the sign $+(-)$ refers to right-handed (left-handed) beam polarization. 
    \label{fig:ee-significance} }
\end{figure}

The rotation angle to the diagonal basis is
\begin{equation}
\tan\xi=\frac{1}{\gamma}\frac{f_V \sth}{f_V\cth \pm f_A\beta},
\end{equation}
for $e^-_R e^+_L$ and  $e^-_L e^+_R$ collisions, respectively, which reproduces the maximally-correlated basis studied in Ref.~\cite{Parke:1996pr}.  We plot the angle relations in  Fig.~\ref{fig:comparebasisee}(a) in the $\xi$-$\theta$ plane.  Similar to $q\bar q\to t\bar t$,  the helicity basis or the rotated beam basis aligns with the optimal (diagonal) basis in the boosted region or near threshold, respectively.  The difference for the electroweak processes is that the spin correlation of $t\bar t$ produced from $e^+e^-$ annihilation is not symmetric under $\theta\to \pi-\theta$, and the $t\bar t$ are mainly entangled in the backward region with $\cos\theta<0$; see Fig.~\ref{fig:fvfa}(b).  Figure~\ref{fig:comparebasisee}(b) shows the Bell inequality violation of the state in angular averaged in different bases versus the boost factor. We see that the optimal (diagonal) basis also leads to a considerable increase in Bell inequality violation in the boosted region.  
To provide a quantitative measurement of the Bell inequality violation, we define a ``significance'' as the ratio of the deviation and its error
\begin{equation}
    \mathcal{S} = \frac{\mathcal{B}-2 }{ \delta\mathcal{B} }, 
\end{equation}
where the uncertainty $\delta\mathcal{B}$ is estimated from the statistical uncertainty of correlation matrix $-$ see e.g.~Ref.~\cite{Han:2023fci} for detail.  We found that the optimal basis choice greatly increases the significance of measuring Bell inequality violation.  We illustrate this for a 1 TeV lepton collider as shown in Fig.~\ref{fig:ee-significance}(a).  Although the $t\bar t$ spin correlation is not sensitive to initial state beam polarization, it is more advantageous to measure Bell inequality violation with a left-hand polarized electron beam and right-hand polarized positron, due to a larger production cross section, as shown in Fig.~\ref{fig:ee-significance}(b). 

The $\gamma\gamma\to t\bar t$ process becomes comparable to $q\bar{q} \to t\bar{t}$ around 4 TeV~\cite{Han:2020uid}, and can be dominant at higher energies. As shown in Appendix~\ref{app:ggttaatt},  this process produces exactly the same spin state and correlations as the $gg\to t\bar t$ process after summing over the color indices, even though their interactions and Feynman diagrams differ. Therefore, the optimal basis choice for $\gamma\gamma\to t\bar t$ process is exactly the same as the basis of $gg\to t\bar t$ process described in Sec.~\ref{sec:qcd}, and the Bell inequality violation of $\gamma\gamma\to t\bar t$ produces nearly the same results as in Fig.~\ref{fig:comparebasisgg}.   Unlike at the LHC where one cannot separate the $t\bar t$ events produced from $q\bar q$ and $gg$, another advantage of lepton colliders is that the annihilation process $e^+e^-\to t\bar t$ and the fusion process $\gamma\gamma\to t\bar t$ can be effectively separated by the invariant mass $m_{t\bar t}$, making it possible to measure the spin singlet Bell state at high-energy lepton colliders.

%%%%%%%%%%%%%%%%%%%%%%%%%%%%%%%%%%%%%%%%%%%%%%%%%%%%%%%%%%%
\section{Conclusion and Discussion}
\label{sec:conclusion}

Studying the entanglement and Bell inequality violation of $t\bar t$ requires interpreting $t\bar t$ as a bipartite quantum state.  For each physical quantum state, quantum entanglement is an intrinsic property and intuitively does not depend on the quantization axis of spin measurements.  In realistic experiments at colliders, however, $t\bar t$ events are analyzed at different phase space points in different Lorentz frames and coordinates.  This situation means that event-dependent bases are often more useful, however, their use yields a fictitious state instead of a genuine quantum state.  The entanglement and Bell inequality violation of fictitious state are no longer an intrinsic property of $t\bar t$ produced at colliders, and they depend on the basis choice of the measurements.

In this work, we point out that the usage of fictitious states is not only valuable to determine the existence of the entanglement and Bell inequality violation, but also advantageous because it allows for an optimization procedure.  We showed that the basis-dependence allows us to find an optimal basis, such that the fictitious state averaged in this basis has the largest signal of entanglement or Bell inequality violation.  By proving that the diagonal basis maximizes both entanglement and Bell inequality violation of an angular-averaged state, we gave the optimal basis choice to study the $t\bar t$ system at colliders.  We summarize our findings as follows:

\begin{itemize}
\item 
We derived an optimal basis choice on an event-by-event basis for a given process, and found that it is approximately the same as the fixed beam basis or the rotated beam basis near the $t\bar t$ production threshold, while it approaches the helicity basis far above threshold. 

\item
For $t\bar t$ events produced from purely $s$-channel processes, namely $e^+e^-\to t\bar t$ and $q\bar q \to t\bar t$, the entanglement and Bell inequality violation are very sensitive to the choice of basis.  Using the optimal basis is especially crucial.  For instance, at a $1\ \tev\ e^+e^-$  collider, the signal of Bell inequality violation in the optimal basis is 40\% larger than the signal of Bell inequality violation in the helicity basis.

\item 
At high-energy hadron colliders, the top pairs are dominantly produced from the $gg\to t\bar t$ process.  The detection of Bell inequality violation at the LHC is very challenging because of the rather weak signal near the $t\bar t$ threshold and the low event rate far above threshold. The situation is improved at higher energies such as at the FCC-hh.

\item 
Near the $t\bar t$ production threshold at the LHC, using the optimal basis leads to a considerable increase in the signal of Bell inequality violation.  In the high-$p_T$ region at the LHC,  \eg~$m_{tt}>1.2$ TeV and $\cos\theta_t < 0.5$, the improvement of Bell inequality violation measurement from the commonly-used helicity basis to the optimal basis is 30\%. Realistic simulations and optimization would be needed to quantify the observability. 
\end{itemize}

We would like to reiterate that, although the presentation in this work is for the $t\bar t$ events, our method and general approach are also applicable to other systems.
Given a quantity calculated from a quantum state, an optimized fictitious state which maximizes the value of the specified quantity can be found following our procedure, even including interactions beyond the Standard Model. This is invaluable as part of the program of quantum tomography for a vast variety of quantum states produced in collider experiments. We anticipate that the program will lead us to a deeper understanding of the quantum world.

\begin{acknowledgments}
We thank Yoav Afik, Alan Barr, Juan de Nova, and Arthur Wu for discussions.  This work was supported in part by the U.S.~Department of Energy under grant No.~DE-SC0007914 and in part by the Pitt PACC.  TH would like to thank the Aspen Center for Physics, where part of this work is complete, which is supported by the National Science Foundation (NSF) grant PHY-1607611.  ML is also supported by the National Science Foundation under grant No.~PHY-2112829. KC is supported in part by the National Science Foundation of China under grant No.~12235001.
\end{acknowledgments}

%%%%%%%%%%%%%%%%%%%%%%%%%%%%%%%%%%%%%%%%%%%%%%%%%%%%%%%%%%%
\appendix
\section{Quantum States at Colliders and Basis Transformations}\label{app:basistransform}

\subsection{Physical Spin State}

In this appendix, we define a physical spin density matrix of quantum states at colliders and discuss basis transformations of the spin density matrix.  We consider the process $I\to t\bar{t}$ where the initial state $I$ is $e^+e^-$, $q\bar q$, or $gg$.  In the center-of-mass frame of $t\bar{t}$, the momentum of $t$ and $\bar{t}$ are
\begin{equation}
p_t^\mu=(E,|p|\mathbf{k}),
\quad\quad\quad
p_{\bar t}^\mu=(E,-|p|\mathbf{k}),
\end{equation}
respectively, where $ \mathbf{k}=(\sin\theta\cos\phi,\sin\theta\sin\phi,\cos\theta)$ is the normalized direction of the momentum of $t$.

The total quantum state $\ket{t\bar{t}}$ from the process $I\to t\bar{t}$ can be expanded in momentum and spin (and color) eigenstates $\ket{\mathbf{k},\alpha\bar{\alpha}}$.  Here, the bipartite particle state $\ket{\mathbf{k},\alpha\bar{\alpha}}$ is defined as $\ket{\mathbf{k},\alpha\bar{\alpha}}=\ket{\mathbf{k},\alpha}\otimes\ket{-\mathbf{k},\bar\alpha}$, where $\ket{\mathbf{k},\alpha}$ denotes the top quark state with spin $\alpha$ and momentum $\mathbf{k}$, and $\ket{-\mathbf{k},\bar\alpha}$ denotes the anti-top quark state with spin $\bar\alpha$ and momentum $-\mathbf{k}$.  The quantum state $\ket{t\bar t}$ expanded in terms of $\ket{\mathbf{k},\alpha\bar\alpha}$ is
\begin{equation}
\ket{t\bar{t}} \propto \int \dk \sum_{\alpha\bar\alpha}\ket{\mathbf{k},\alpha\bar{\alpha}}\bra{\mathbf{k},\alpha\bar{\alpha}}T\ket{I,\lambda}=\int \dk \sum_{\alpha\bar\alpha} \mathcal{M}^{\lambda}_{\alpha\bar{\alpha}}(\mathbf{k})\ket{\mathbf{k},\alpha\bar{\alpha}},
\end{equation}
where $\lambda$ denotes the quantum number of the initial state, $T$ is the transition matrix and $\mathcal{M}^\lambda_{\alpha\bar\alpha}(\mathbf{k})$ is the corresponding amplitude.  Typically, the initial state is a mixed state, \ie, $\rho^I =\sum_{\lambda,\lambda'}\rho^I_{\lambda\lambda'}\ket{I,\lambda}\bra{I,\lambda'}$.

The production density matrix is
\begin{equation}\label{eq:Rcomplete}
    \hat\rho^{tt}\propto\int\dk\dk' \sum_{\alpha\bar\alpha}\sum_{\alpha'\bar\alpha'}
    \mathcal{M}^{\lambda}_{\alpha\bar{\alpha}}(\mathbf{k})\rho^I_{\lambda\lambda'}
    (\mathcal{M}^{\lambda'}_{\alpha'\bar{\alpha}'}(\mathbf{k}'))^*
    \ket{\mathbf{k},\alpha\bar{\alpha}}\bra{\mathbf{k}',\alpha'\bar{\alpha}'},
\end{equation}
and this complete density matrix is defined in spin space, color space,\footnote{The color index is omitted here for simplicity.} and momentum space, but we only focus on the entanglement in spin space.  To obtain a spin density matrix, we first need to trace over the state in the color space, then we can either project the state into an eigenstate in the momentum space or trace the state over the momentum space. 

By projecting the quantum state into a momentum eigenstate, \ie, selecting the $t\bar{t}$ with fixed scattering angle, we obtain a quantum sub-state for $t\bar{t}$ produced at a fixed phase space point
\begin{equation}
\label{eq:Rk}
\hat\rho(\mathbf{k})=\bra{\mathbf{k}} \hat\rho \ket{\mathbf{k}} \nn 
=\rho(\mathbf{k})_{\alpha\alpha',\bar\alpha\bar\alpha'} \ket{\alpha\bar{\alpha}}\bra{\alpha'\bar{\alpha}'},
\end{equation}
where $\rho(\mathbf{k})_{\alpha\bar\alpha,\alpha'\bar\alpha'}$ is the density matrix of the state $\hat\rho(\mathbf{k})$ expressed in the basis $\ket{\alpha \bar\alpha}$. 

By tracing the quantum state over the momentum space, we obtain the angular average of $\rho(\mathbf{k})$ in a phase space region $\Pi$,
\begin{equation}
\label{eq:physicalrho}
\begin{aligned}
    \hat\rho_{\Pi}&=\tr_{\mathbf{k}\in\Pi}\Big(\hat\rho\ket{\mathbf{k}}\bra{\mathbf{k}}\Big)
    &=\frac{1}{\sigma_\Pi}\int_{\Omega\in\Pi}\dO \frac{\rm d\sigma}{\dO} \rho(\mathbf{k})_{\alpha\alpha',\bar\alpha\bar\alpha'} \ket{\alpha\bar{\alpha}}\bra{\alpha'\bar{\alpha}'}.
\end{aligned}
\end{equation}
Here, $\sigma_\Pi=\int_{\Omega\in\Pi}\dO \frac{\rm d\sigma}{\dO}$ is the integrated cross section in the region $\Omega\in\Pi$.  To obtain the density matrix for the physical state $\hat \rho_\Pi$, we need to average the density matrix $\rho(\mathbf{k})_{\alpha\bar\alpha,\alpha'\bar\alpha'}$ of the quantum sub-states in an event-independent basis.

%%%%%%%%%%%%%%%%
\subsection{Transformations between Spin Bases}
\label{sub:basisrotation}

To represent the spin density matrix $\rho$ with a $4\times 4$ matrix $\rho_{\alpha\alpha',\bar\alpha\bar\alpha'}$, we need to choose a basis for the spin measurements.  The bases of spin eigenstates $\ket{\alpha\bar{\alpha}}= \ket{\uparrow\uparrow},\ket{\uparrow\downarrow},\ket{\downarrow\uparrow},\ket{\downarrow\downarrow}$ can be defined based on the quantization axes $\vec{e}_i$.  Specifically, $\ket{\uparrow}$ or $\ket{\downarrow}$ denote the state with positive or negative spin along the third direction $\vec{e}_3$ in the rest frame of (anti-)top quark, and we have
\begin{equation}\label{eq:appendixbasis1}
\begin{split}
\vec{\sigma}\cdot \vec{e}_1&=\ket{\uparrow}\bra{\downarrow}+\ket{\downarrow}\bra{\uparrow},\\
\vec{\sigma}\cdot \vec{e}_2&=-i\ket{\uparrow}\bra{\downarrow}+i\ket{\downarrow}\bra{\uparrow},\\
\vec{\sigma}\cdot \vec{e}_3&=\ket{\uparrow}\bra{\uparrow}-\ket{\uparrow}\bra{\uparrow},
\end{split}
\end{equation}
with $\vec{\sigma}\cdot \vec{e}_3 \ket{\uparrow}=\ket{\uparrow}, \vec{\sigma}\cdot \vec{e}_3 \ket{\downarrow}=-\ket{\downarrow}$.  A different choice of quantization axis yields the spin eigenstates
\begin{equation}\label{eq:appendixbasis2}
\begin{split}
\vec{\sigma}\cdot \vec{e}_1'&=\ket{\uparrow}'\bra{\downarrow}'+\ket{\downarrow}'\bra{\uparrow}',\\
\vec{\sigma}\cdot \vec{e}_2'&=-i\ket{\uparrow}'\bra{\downarrow}'+i\ket{\downarrow}'\bra{\uparrow}',\\
\vec{\sigma}\cdot \vec{e}_3'&=\ket{\uparrow}'\bra{\uparrow}'-\ket{\uparrow}'\bra{\uparrow}'.
\end{split}
\end{equation}
Here, two different quantization axis choices $\vec{e}_i$ and $\vec{e}_i'$ can be related by an SO(3) rotation
\begin{equation}
    \vec{e}_i' = R_{ji} \vec{e}_j,
\end{equation}
and two spin bases are related by an SU(2) transformation
\begin{equation}\label{eq:appendixSU2}
\ket{\beta}'=U_{\alpha\beta} \ket{\alpha}.
\end{equation}
Substituting Eq.~\eqref{eq:appendixSU2} into Eq.~\eqref{eq:appendixbasis2} and we obtain
\begin{equation}
    R_{ij}=\frac{1}{2}\tr(U^\dagger \sigma_i U \sigma_j).
\end{equation}
We can also express the explicit form of $R_{ij}$ and $U_{\alpha\beta}$ in terms of Euler angles. The spin bases and quantization axes shown in Fig.~\ref{fig:PlotCoordinates} are related by
\begin{subequations}\label{eq:R-EulerAngle}
\begin{align}
\vec{e}_i^{\rm rotated}&= R(0,\phi)_{ji} \vec{e}_j^{\rm fixed}, \\
\vec{e}_i^{\rm helicity}&= R(\theta,\phi)_{ji} \vec{e}_j^{\rm fixed}, \\
\vec{e}_i^{\rm diag}&= R(\theta-\xi,\phi)_{ji} \vec{e}_j^{\rm fixed},
\end{align}
\end{subequations}
and
\begin{subequations}
\label{eq:U-EulerAngle}
\begin{align}
\ket{\alpha\bar{\alpha}}^{\rm rotated}&=U(0,\phi)_{\beta\alpha,\bar\beta\bar\alpha}\ket{\beta\bar\beta}^{\rm fixed},\\
\ket{\alpha\bar{\alpha}}^{\rm helicity}&=U(\theta,\phi)_{\beta\alpha,\bar\beta,\bar\alpha}\ket{\beta\bar\beta}^{\rm fixed},\\
\ket{\alpha\bar{\alpha}}^{\rm diag}&=U(\theta-\xi,\phi)_{\beta\alpha,\bar\beta\bar\alpha}\ket{\beta\bar\beta}^{\rm fixed},
\end{align}
\end{subequations}
where $R(\theta,\phi)=e^{-i\theta S_y}e^{-i\phi S_z}$, $U_{\beta\alpha,\bar\beta\bar\alpha}=U_{\beta\alpha}\otimes U_{\bar\beta\bar\alpha}$, and $U(\theta,\phi)\equiv\left(e^{-i\theta\frac{\sigma_y}{2}}e^{-i\phi\frac{\sigma_z}{2}} \right)\otimes \left(e^{-i\theta\frac{\sigma_y}{2}}e^{-i\phi\frac{\sigma_z}{2}} \right)$.

%%%%%%%%%%%%%%%%%%%%%%%%%%%%%%%%%%%%%%%%%%%%%%%%%%%%%%%%%%%
\section{Entanglement of Fictitious States}
\label{app:fictitiousstate}

In this section, we prove that a fictitious state with non-zero entanglement implies that there exists a quantum sub-state with non-zero entanglement.  The convex sum of quantum sub-states in an event-dependent basis gives a fictitious state,
\begin{align}
 \bar\rho_{\alpha\beta,\bar\alpha\bar\beta}^{\rm fic} &=\frac{1}{\sigma_\Pi}\int_{\Omega\in \Pi} \dO \frac{\rm d\sigma}{\dO}  \rho(\mathbf{k})^{}_{\alpha\beta,\bar\alpha\bar{\beta}}.
\end{align}
The spin correlation matrix of the fictitious state is also a convex sum of the spin correlation matrix of quantum sub-states
\begin{align}
 \bar C^{\rm fic}_{ij} &=\frac{1}{\sigma_\Pi}\int_{\Omega\in \Pi} \dO \frac{\rm d\sigma}{\dO} C(\mathbf{k})_{ij}.
\end{align}
We start from concurrence. From Eq.~\eqref{eq:concurrence-lambda1234} it is difficult to see the relationship between the concurrence of quantum sub-states and the concurrence of their angular average, so we turn to the separability condition.  For bipartite qubit systems, $\mathscr{C}(\rho)=0$ is a necessary and sufficient condition of separability~\cite{Wootters:1997id}. A quantum state of $t\bar t$ is separable if it can be written as a convex sum of $\rho^t\otimes \rho^{\bar t}$ as follows,
\begin{equation}
    \rho=\sum_{i} p_i \rho^t_{i}\otimes \rho^{\bar t}_i.
\end{equation}
If all the quantum sub-states $\rho(\mathbf{k})$ are separable, then their angular average $\bar\rho$, a convex sum of $\rho(\mathbf{k})$, is obviously separable. Therefore, the entanglement in the fictitious states implies the existence of entanglement in a quantum sub-state.

The violation of Bell inequality is similar. If all the quantum sub-states satisfy Bell's inequality Eq.~\eqref{eq:Bella1a2b1b2} for any four spatial directions $\vec{a}_1,\vec{a}_2,\vec{b}_1$, and $\vec{b}_2$, \ie,
\begin{align}\label{eq:CHSHforfixK}
    \Vec{a}_1\cdot C(\mathbf{k})(\vec{b}_1-\Vec{b}_2)+
    \Vec{a}_2\cdot C(\mathbf{k})(\vec{b}_1+\Vec{b}_2) \in [-2,2],
\end{align}
then the angular average (convex sum) of Eq.~\eqref{eq:CHSHforfixK} gives
\begin{align}
    &\frac{1}{\sigma_\Pi} \int_{\Omega\in\Pi} \dO \frac{\rm d\sigma}{\dO}
    \left( \Vec{a}_1\cdot C(\mathbf{k})(\vec{b}_1-\Vec{b}_2)+ \Vec{a}_2\cdot C(\mathbf{k})(\vec{b}_1+\Vec{b}_2) \right)\nn
    =&~ \Vec{a}_1\cdot \bar C \cdot(\vec{b}_1-\Vec{b}_2)+ \Vec{a}_2\cdot \bar C \cdot(\vec{b}_1+\Vec{b}_2) \\
    \in&~  [-2,2],
\end{align}
and the fictitious state also satisfies Bell's inequality. Therefore, if the violation of Bell's inequality is observed in the fictitious state, there must be some quantum sub-state violating Bell's inequality.

%%%%%%%%%%%%%%%%%%%%%%%%%%%%%%%%%%%%%%%%%%%%%%%%%%%%%%%%%%%
\section{Generalized Proof of the Optimal Basis}\label{app:proof}

We start with an arbitrary event-dependent basis choice.  The spin correlation matrix in this basis is
\begin{equation}
\bar C_{ij} =\intk C_{ij,\Omega}.
\end{equation}
When $\bar C_{ij}$ is not symmetric, the optimal Bell inequality violation is related to the singular values of $C_{ij}$ rather than the eigenvalues of $C_{ij}$.  Unlike in the main text, in this appendix we use $\bar c_i$ to denote the singular value of $\bar C_{ij}$,
\begin{equation}
(\bar c_1, \bar c_2, \bar c_3) = \operatorname{singular-value} \left[\bar C^{\rm basis}\right].
\end{equation}
The singular values are all positive. Without loss of generality, we assume $\bar c_1 \geq \bar c_2 \geq \bar c_3$.

Similarly, in this appendix we use $\mu_{\Omega}$ to denote the singular values instead of the eigenvalues of a quantum sub-state.  The diagonal basis is obtained from an event-dependent singular value decomposition:
\begin{equation}%\label{eq:Csingular}
    \bfC^{\rm diag}_\Omega=R_{\Omega}^{t}\bfC_\Omega \left(R^{\bar t}_{\Omega}\right)^T=\begin{pmatrix}
        \mu_{1,\Omega} &0&0\\
        0&\mu_{2,\Omega} &0\\
        0&0&\mu_{3,\Omega} \\
    \end{pmatrix}.
\end{equation}
In the diagonal basis, the singular values of the angular-averaged spin correlation matrix are
\begin{equation}
    \bar c^{\rm diag}_i = \intk \mu_{i,\Omega} \equiv \bar \mu_i.
\end{equation}
In Eqs.~\eqref{eq:trC1trC2} and~\eqref{eq:mu1Ciimu3}, we showed that the eigenvalues of a general angular-averaged spin correlation matrix are bounded by the eigenvalues of $\bar C^{\rm diag}$.  Similarly, the singular values of the angular-averaged spin correlation matrix $\bar C$ in a general basis, namely $\bar c_i$, satisfy the following relations
\begin{align}
\label{eq:cileqmu1}
    &\bar c_i \leq \bar \mu_{1},\\
\label{eq:c12leqmu12}
    &\bar c_1 + \bar c_2 \leq \bar \mu_1 + \bar \mu_2 .
\end{align}
Eq.~\eqref{eq:cileqmu1} holds because an entry of a matrix must be smaller than its largest singular value.  To prove Eq.~\eqref{eq:c12leqmu12}, we start with two lemmas:

\vspace{10pt}
\noindent{\bf Lemma 1}: The trace of any real matrix $\mathbf{C}$ is smaller than the sum of its singular values. \\
\vspace{5pt}
\noindent{\it Proof:} The singular value decomposition is $\bfC^{\rm diag}=R_l \bfC R_r^T=\operatorname{diag}(\mu_1,\cdots \mu_n)$, so $\tr(\bfC)=\tr(R_l^T \bfC^{\rm diag} R_r)=\tr(R^{lr} \bfC^{\rm diag})= \sum_{i=1}^n R^{lr}_{ii} \mu_i $, where $R^{lr}\equiv R_r^T R_l^T$.  Because $R^{lr}$ is an orthogonal rotation, all of its elements are smaller than one, so $\tr(\bfC)= \sum_{i=1}^n R^{lr}_{ii} \mu_i \leq \sum_{i=1}^n  \mu_i $

\vspace{10pt}
\noindent{\bf Lemma 2}: Consider a general matrix $\mathbf{A}$ and the first $2\times 2$ submatrix $\mathbf{A}_{2\times 2}$ of it, the sum of its largest two singular values is always larger than or equal to the sum of the singular values of $\mathbf{A}_{2\times 2}$. \\
\vspace{5pt} \noindent{\it Proof:} see Ref.~\cite{THOMPSON19721}.
\\ 
\vspace{10pt}

Next, we denote three left-singular (right-singular) vectors of $\bar C$ as $\vec u_i$ ($\vec v_i$).  The corresponding singular value is then
\begin{align}
    \bar c_i=\vec{u}_i\cdot \overline \bfC \cdot \vec{v}_i
    =\intk \left( \vec{u}_i\cdot \bfC_\Omega \cdot \vec{v}_i \right) \equiv \intk c_{i,\Omega},
\end{align}
where we have defined $c_{i,\Omega}=\vec{u}_i\cdot \bfC_\Omega \cdot \vec{v}_i$, so that $\bar c_i$ is a convex sum of $c_{i,\Omega}$.  In addition, we define another matrix $C'_{\Omega}$ as
\begin{equation}
    C'_{ij,\Omega} = \vec{u}_i\cdot \bfC_\Omega \cdot \vec{v}_j.
\end{equation}
The three singular value of $\bfC'_{\Omega}$ are $\mu_{i,\Omega}$, and the trace of first $2\times2$ submatrix of $\bfC'_{\Omega}$ is  $c_{1,\Omega}+c_{2,\Omega}$. 
From the above two lemmas, we find that $c_{1,\Omega}+c_{2,\Omega} \leq
\mu_{1,\Omega}+\mu_{2,\Omega}$, which shows Eq.~\eqref{eq:c12leqmu12}. 

With Eqs.~\eqref{eq:cileqmu1} and \eqref{eq:c12leqmu12}, we are ready to show that $\bar c_1^2 + \bar c_2^2 < \bar \mu_1^2 + \bar \mu_2^2$.  The proof is nearly the same with Case (a) in Sec.~\ref{sub:Bellinequalities}.

If $0\leq \bar c_1 \leq \bar \mu_2$, we have $\bar c_1^2 + \bar c_2^2 \leq 2\bar\mu_2^2 \leq \bar\mu_1^2+\bar\mu_2^2 $ directly.

If $\bar \mu_2\leq \bar c_1 \leq\bar \mu_1$, we first have
\begin{align}
\label{eq:mu2cimu1:app}
-\frac{\bar \mu_1-\bar \mu_2}{2}
\leq\frac{\bar \mu_1+\bar\mu_2}{2}-\bar c_1 \leq & \frac{\bar\mu_1-\bar\mu_2}{2}.
\end{align}
Then, according to $\bar c_1+\bar c_2 \leq \bar \mu_1+ \bar \mu_2$, we have,
\begin{equation}
\label{eq:maximumofCiiCjj2:app}
\bar c_1^2 + \bar c_2^2 \leq \bar c_1^2 + \left(\bar \mu_1 + \bar \mu_2 - \bar c_1\right)^2.
\end{equation}
Next, we define the same function as in Sec.~\ref{sub:Bellinequalities}:
\begin{align}
f(\Delta) &= \left(\frac{\bar \mu_1+\bar \mu_2}{2}+\Delta\right)^2+ \left(\frac{\bar \mu_1+\bar \mu_2}{2}-\Delta\right)^2, \nn
&=\frac{(\bar \mu_1+\bar \mu_2)^2}{2}+2\Delta^2,
\end{align}
which satisfies $f(\Delta_1)\leq f(\Delta_2)$ for $|\Delta_1|\leq |\Delta_2|$.  Then $\bar\mu_1^2+\bar \mu_2^2$ and the right-hand side of Eq.~\eqref{eq:maximumofCiiCjj2:app} and can be rewritten using $f(\Delta)$:
\begin{align}
\label{eq:fmu:app}
 \bar c_i^2+ \left(\bar \mu_1+\bar \mu_2-\bar c_i\right)^2&=f\left(\frac{\bar\mu_1+\bar\mu_2}{2}-\bar c_i \right),\\
\label{eq:fmu2:app}
\bar\mu_1^2+\bar\mu_2^2&= f\left(\frac{\bar\mu_1-\bar\mu_2}{2}\right).
\end{align}
From Eq.~\eqref{eq:mu2cimu1:app}, the right-hand side of Eq.~\eqref{eq:fmu:app} is smaller than the right-hand side of Eq.~\eqref{eq:fmu2:app}, and it follows that $\bar c_i^2 + \left(\bar \mu_1 + \bar \mu_2 - \bar c_i\right)^2 \leq \bar\mu_1^2+\bar\mu_2^2$.  In combination with Eq.~\eqref{eq:maximumofCiiCjj2:app}, we reach our conclusion, $\bar c_i^2+ \bar c_j^2 \leq \bar \mu_1^2 + \bar \mu_2^2 $.

%%%%%%%%%%%%%%%%%%%%%%%%%%%%%%%%%%%%%%%%%%%%%%%%%%%%%%%%%%%
\section{Angular Momentum Conservation}
\label{app:LzConservation}

Unlike with a general density matrix, as parameterized by Eq.~\eqref{eq:rhoBiCij}, the quantum states produced at a collider are constrained by the angular momentum conservation of the scattering processes.  This leads to some consequences closely related to the entanglement of $t\bar t$ system.

Denote the total spin of the initial states along the beamline as $S_z^I$, then the $t\bar t$ state produced at the collider must have the following form
\begin{equation}\label{eq:amplitudeSz}
    \ket{t\bar t}\propto e^{iS_z^I\phi}\left( e^{-i\phi}\mathcal{M}_{\uparrow\uparrow}\ket{\uparrow\uparrow} + \mathcal{M}_{\uparrow\downarrow}\ket{\uparrow\downarrow}+\mathcal{M}_{\downarrow\uparrow}\ket{\downarrow\uparrow}+ e^{i\phi}\mathcal{M}_{\downarrow\downarrow}\ket{\downarrow\downarrow} \right),
\end{equation}
in the fixed beam basis. 

At leading order, the polarized amplitudes $\mathcal{M}_{\alpha\bar\alpha}$ are real (with respect to a common phase), then the correlation matrix in the fixed beam basis is
\begin{equation}
C^{\rm fixed}\propto
\scalemath{0.8}{
    \begin{pmatrix}
        2 \left(\mathcal{M}_{\downarrow \downarrow } \mathcal{M}_{\uparrow \uparrow } \cos (2 \phi
   )+\mathcal{M}_{\downarrow \uparrow } \mathcal{M}_{\uparrow \downarrow }\right) & 2 \mathcal{M}_{\downarrow
   \downarrow } \mathcal{M}_{\uparrow \uparrow } \sin (2 \phi ) & 2 \cos (\phi ) \left(\mathcal{M}_{\downarrow
   \uparrow } \mathcal{M}_{\uparrow \uparrow }-\mathcal{M}_{\downarrow \downarrow } \mathcal{M}_{\uparrow
   \downarrow }\right) \\
 2 \mathcal{M}_{\downarrow \downarrow } \mathcal{M}_{\uparrow \uparrow } \sin (2 \phi ) & 2
   \mathcal{M}_{\downarrow \uparrow } \mathcal{M}_{\uparrow \downarrow }-2 \mathcal{M}_{\downarrow \downarrow }
   \mathcal{M}_{\uparrow \uparrow } \cos (2 \phi ) & 2 \sin (\phi ) \left(\mathcal{M}_{\downarrow \uparrow }
   \mathcal{M}_{\uparrow \uparrow }-\mathcal{M}_{\downarrow \downarrow } \mathcal{M}_{\uparrow \downarrow
   }\right) \\
 2 \cos (\phi ) \left(\mathcal{M}_{\uparrow \downarrow } \mathcal{M}_{\uparrow \uparrow
   }-\mathcal{M}_{\downarrow \downarrow } \mathcal{M}_{\downarrow \uparrow }\right) & 2 \sin (\phi )
   \left(\mathcal{M}_{\uparrow \downarrow } \mathcal{M}_{\uparrow \uparrow }-\mathcal{M}_{\downarrow \downarrow
   } \mathcal{M}_{\downarrow \uparrow }\right) & \mathcal{M}_{\downarrow \downarrow }^2-\mathcal{M}_{\downarrow
   \uparrow }^2-\mathcal{M}_{\uparrow \downarrow }^2+\mathcal{M}_{\uparrow \uparrow }^2 \\
    \end{pmatrix}.
    }
\end{equation}
The spin correlation matrix in the rotated beam basis is
\begin{equation}
C^{\rm rotated}\propto
    \begin{pmatrix}
        2 \mathcal{M}_{\downarrow \downarrow } \mathcal{M}_{\uparrow \uparrow }+2 \mathcal{M}_{\downarrow \uparrow }
   \mathcal{M}_{\uparrow \downarrow } & 0 & 2 \mathcal{M}_{\downarrow \uparrow } \mathcal{M}_{\uparrow \uparrow
   }-2 \mathcal{M}_{\downarrow \downarrow } \mathcal{M}_{\uparrow \downarrow } \\
 0 & 2 \mathcal{M}_{\downarrow \uparrow } \mathcal{M}_{\uparrow \downarrow }-2 \mathcal{M}_{\downarrow
   \downarrow } \mathcal{M}_{\uparrow \uparrow } & 0 \\
 2 \mathcal{M}_{\uparrow \downarrow } \mathcal{M}_{\uparrow \uparrow }-2 \mathcal{M}_{\downarrow \downarrow }
   \mathcal{M}_{\downarrow \uparrow } & 0 & \mathcal{M}_{\downarrow \downarrow }^2-\mathcal{M}_{\downarrow
   \uparrow }^2-\mathcal{M}_{\uparrow \downarrow }^2+\mathcal{M}_{\uparrow \uparrow }^2 \\
    \end{pmatrix}.
\end{equation}
As a consequence of Eq.~\eqref{eq:amplitudeSz}, the spin correlation matrix of $t\bar t$ is diagonal in the normal direction of the scattering plane.

Another consequence of angular momentum conservation is that the azimuthal angular average in the fixed beam basis usually leads to a complete cancellation of entanglement. Assume that we have produced a maximally entangled state $\ket{\Psi_i}$ (\ie, a Bell state) at $\phi=0$,
\begin{align}
    \ket{\Psi_0}&= \frac{\ket{\uparrow\downarrow}-\ket{\downarrow\uparrow}}{\sqrt{2}}, \\
    \ket{\Psi_1}&= \frac{\ket{\uparrow\uparrow}-\ket{\downarrow\downarrow}}{\sqrt{2}}, \\
    \ket{\Psi_2}&=i \frac{\ket{\uparrow\uparrow}+\ket{\downarrow\downarrow}}{\sqrt{2}}, \\
    \ket{\Psi_3}&=- \frac{\ket{\uparrow\downarrow}+\ket{\downarrow\uparrow}}{\sqrt{2}}.
\end{align}
Here, $\ket{\Psi_0}$ is a spin singlet state and the other three comprise a spin triplet. Then we must have
\begin{align}
    \ket{\Psi_0(\phi)}&= \frac{\ket{\uparrow\downarrow}-\ket{\downarrow\uparrow}}{\sqrt{2}}, \\
    \ket{\Psi_1(\phi)}&= \frac{\ket{\uparrow\uparrow}-e^{2i\phi}\ket{\downarrow\downarrow}}{\sqrt{2}}, \\
    \ket{\Psi_2(\phi)}&= i \frac{\ket{\uparrow\uparrow}+e^{2i\phi}\ket{\downarrow\downarrow}}{\sqrt{2}}, \\
    \ket{\Psi_3(\phi)}&=- \frac{\ket{\uparrow\downarrow}+\ket{\downarrow\uparrow}}{\sqrt{2}}
\end{align}
away from $\phi=0$ (with respect to a global phase).  After integrating the density matrix $\int d\phi \ket{\Psi_i(\phi)}\bra{\Psi_i(\phi)}$ over the full range of $\phi$, we cannot observe entanglement for $\ket{\Psi_1}$ and $\ket{\Psi_2}$.

In the high-$p_T$ region for most SM processes, the entanglement of $t\bar t$ is large from the triplet component $\ket{\Psi_2}$.  Therefore for the quantum state, which is necessarily averaged in fixed beam basis, 
the azimuthal angular average of the density matrix is
\begin{equation}
  \bar \rho^{\rm fixed}_{\rm triplet}
=\frac{1}{2\pi} \int \dphi \ket{\Psi_2(\phi)}\bra{\Psi_2(\phi)} =\frac{\ket{\uparrow\uparrow}\bra{\uparrow\uparrow}+\ket{\downarrow\downarrow}\bra{\downarrow\downarrow}}{2},
\end{equation}
which is a separable state.

%%%%%%%%%%%%%%%%%%%%%%%%%%%%%%%%%%%%%%%%%%%%%%%%%%%%%%%%%%%
\section{Top Anti-top Produced from Gluon and Photon Scattering}
\label{app:ggttaatt}

In this appendix, we show that $\gamma\gamma \to t\bar t$ produces exactly the same spin density matrix as $gg\to t\bar t$ after summing over the color indices, even though their interactions and Feynman diagrams differ.  The amplitudes of $g_a(p_1)g_b(p_2)\to t_i(p_3)\bar t_j(p_4)$ can be parametrized as
\begin{equation}\label{eq:Amplitudeggtt}
    i\mathcal{M}(g_a g_b \to t_i \bar t_j) = -i g_s^2 \Big[(T^a T^b)_{ij} \Mt + (T^b T^a)_{ij} \Mu + i f^{abc} T^c_{ij} \Ms \Big],
\end{equation}
where $a,b$, and $c$ are the color indexes of gluon, $i$ and $j$ are the color indexes of quarks, $T^a$ are the generators of $SU(N)$, and $f^{abc}$ is the structure constant.  $\mathcal{M}_{s}$, $\mathcal{M}_t$, and $\mathcal{M}_u$ are from the amplitude of $s$-channel, $t$-channel, and $u$-channel diagrams
\begin{align}
    \Ms &= \frac{\bar u_3 \gamma_\rho v_4}{\hat s} \epsilon_{1\mu} \epsilon_{2\mu} \left[(p_1-p_2)^\rho g^{\mu\nu}+(p_2-k)^\mu g^{\nu\rho}+(k-p1)^\mu g^{\nu\rho}\right], \\
    \Mt &= \bar u_3 \frac{\eslash_1 (\pslash_t-m)\eslash_2}{p_t^2-m^2} v_4,\\
    \Mu &= \bar u_3 \frac{\eslash_2 (\pslash_u-m)\eslash_1}{p_u^2-m^2} v_4,
\end{align}
where $k=-p_3-p_4$, $p_t=p_2-p_4$,  $p_u=p_1-p_4$, and $m$ is the mass of the top quark.  The denominator of the propagator written in terms of velocity factor $\beta$ and scattering angle $\theta$ are $p_t^2-m^2=\frac{\hat s}{2}(\beta c_\theta-1)$,  $p_u^2-m^2=-\frac{\hat s}{2}(\beta c_\theta+1)$, and $k^2=\hat s$.

For the QED processes  $\gamma\gamma\to t\bar  t$, the amplitude is simply
\begin{equation}\label{eq:Amplitudeaatt}
i\mathcal{M}(\gamma\gamma\to t\bar  t)= -ie^2 (\Mt+\Mu).
\end{equation}
Before comparing the spin density matrix from Eq.~\eqref{eq:Amplitudeggtt} and Eq.~\eqref{eq:Amplitudeaatt}, we first list some general identities.  Based on $\{\gamma^\mu,\gamma^\nu\} = 2g^{\mu\nu}$ and the Dirac equations of the spinors, we have
\begin{subequations}
\begin{align}
    \bar u_3 \eslash_1 \pslash_3 \eslash_2 v_4 & = 2\bar u_3 \eslash_2 v_4 (p_3\cdot \epsilon_1) - m \bar u_3 \eslash_1 \eslash_2 v_4 ,\\
    \bar u_3 \eslash_1 \pslash_4 \eslash_2 v_4 & = 2\bar u_3 \eslash_1 v_4 (p_4\cdot \epsilon_2) + m \bar u_3 \eslash_1 \eslash_2 v_4 ,\\
    \bar u_3 \eslash_2 \pslash_3 \eslash_1 v_4 & = 2\bar u_3 \eslash_1 v_4 (p_3\cdot \epsilon_2) - m \bar u_3 \eslash_2 \eslash_1 v_4 ,\\
    \bar u_3 \eslash_2 \pslash_4 \eslash_1 v_4 & = 2\bar u_3 \eslash_2 v_4 (p_4\cdot \epsilon_1) + m \bar u_3 \eslash_2 \eslash_1 v_4 ,
\end{align}
\end{subequations}
which lead to
\begin{align}\label{eq:p3p4identity}
    \bar u_3 \eslash_1 (\pslash_3-\pslash_4) \eslash_2 v_4 -& (\epsilon_1 \leftrightarrow \epsilon_2) \nn
    =& 2\bar u_3 \eslash_1 v_4 (k\cdot \epsilon_2) -  2\bar u_3 \eslash_2 v_4 (k\cdot \epsilon_1) - 2 m \bar u_3 [\eslash_1, \eslash_2] v_4.
\end{align}
Based on the transverse conditions $p_1\cdot \epsilon_1=0$ and $p_2\cdot \epsilon_2=0$, we have
\begin{subequations}
    \begin{align}
    \eslash_1\pslash_1\eslash_2 + \eslash_2\pslash_1\eslash_1 & = 2 (p_1\cdot \epsilon_2) \eslash_1 - 2\pslash_1 (\epsilon_1 \cdot \epsilon_2), \\
    \eslash_1\pslash_2\eslash_2 + \eslash_2\pslash_2\eslash_1 & = 2 (p_2\cdot \epsilon_1) \eslash_2 - 2\pslash_2 (\epsilon_1 \cdot \epsilon_2),
    \end{align}
\end{subequations}
which lead to
\begin{align}\label{eq:p1p2idensity}
    \eslash_1 (\pslash_1-\pslash_2) \eslash_2 +& (\epsilon_1 \leftrightarrow \epsilon_2) \nn
    =& 2\eslash_1 (p_1\cdot \epsilon_2) -  2\eslash_2 (p_2\cdot \epsilon_1) - 2 (\pslash_1-\pslash_2) (\epsilon_1\cdot \epsilon_2).
\end{align}
For the QED process, the spin density matrix is directly given by the helicity amplitudes,
\begin{equation}
    \rho^{\rm QED}_{\alpha\bar\alpha,\alpha'\bar\alpha'} \propto \mathcal{M}_{\alpha\bar \alpha}(\gamma\gamma\to t \bar t) \mathcal{M}_{\alpha'\bar \alpha'}^*(\gamma\gamma\to t\bar t).
\end{equation}
For the QCD process, we need to sum over color to obtain the spin density matrix,
\begin{equation}\label{eq:rhoqcd}
    \rho^{\rm QCD}_{\alpha\bar\alpha,\alpha'\bar\alpha'} \propto \sum_{a,b,i,j}
    \mathcal{M}_{\alpha\bar \alpha}(g_a g_b\to t_{i}\bar t_{j}) \mathcal{M}_{\alpha'\bar \alpha'}^*(g_a g_b\to t_{i}\bar t_{j}).
\end{equation}
To compare these two spin density matrices, we decompose the $gg\to t\bar t$ process according to the color configuration,
\begin{align}\label{eq:AmpggttColorFactor}
    \mathcal{M}(g_a g_b \to t_i \bar t_j)  &= g_s^2 \Big[(T^a T^b)_{ij} \Mt + (T^b T^a)_{ij} \Mu + i f^{abc} T^c_{ij} \Ms \Big]\nn
    &= g_s^2 \left[ \{T^a, T^b\}_{ij} \frac{\Mt+\Mu}{2}  + i f^{abc} T^c_{ij} \left(\frac{\Mt-\Mu}{2}+\Ms\right) \right].
\end{align}
Since there is no interference between the gluon color symmetric part and asymmetric part,
\begin{equation}\label{eq:colorsymasym}
    \sum_{a,b,i,j} f^{abc} T^c_{ij} ~ \{T^a, T^b\}_{ij}=0,
\end{equation}
they contribute to the $t\bar t$ spin correlation separately.  The color symmetric contribution is exactly the same as the $\gamma\gamma \to t\bar t$.  

Next, we show that the color asymmetric contribution also produces the same spin density matrix with $\gamma\gamma\to t\bar t$, by proving
\begin{equation}
\frac{\Mt-\Mu}{2}+\Ms \propto \Mt + \Mu.
\end{equation}

By decomposing $\Mt+\Mu$ and $\Mt-\Mu$ into symmetric and asymmetric part with respect to $\epsilon_1$ and $\epsilon_2$, we have
\begin{align}
    \Mt+\Mu =\frac{2}{\hat s (\beta^2 c_\theta^2 -1)} 
    &\Bigg[ \eslash_1 \left(\frac{\pslash_t+\pslash_u+2m}{2} + \frac{\pslash_t-\pslash_u}{2} \beta c_\theta \right)\eslash_2 + (\epsilon_1 \leftrightarrow \epsilon_2) \nn
    &+ \eslash_1 \left(\frac{\pslash_t+\pslash_u+2m}{2}\beta c_\theta + \frac{\pslash_t-\pslash_u}{2}\right)\eslash_2 - (\epsilon_1 \leftrightarrow \epsilon_2)
    \Bigg],\\
    \Mt-\Mu =\frac{2}{\hat s (\beta^2 c_\theta^2 -1)} 
    &\Bigg[ \eslash_1 \left(\frac{\pslash_t+\pslash_u+2m}{2}\beta c_\theta + \frac{\pslash_t-\pslash_u}{2}  \right)\eslash_2 + (\epsilon_1 \leftrightarrow \epsilon_2) \nn
    &+ \eslash_1 \left(\frac{\pslash_t+\pslash_u+2m}{2}+ \frac{\pslash_t-\pslash_u}{2}\beta c_\theta \right)\eslash_2 - (\epsilon_1 \leftrightarrow \epsilon_2)
    \Bigg].
\end{align}
Then, we have
\begin{align}
    \beta c_\theta (\Mt+\Mu) - (\Mt-\Mu)
    =& \frac{2}{\hat s} \bar u_3 \Bigg[ \eslash_1 \left(\frac{\pslash_t-\pslash_u}{2} \right)\eslash_2 + \eslash_2 \left(\frac{\pslash_t-\pslash_u}{2} \right)\eslash_1  \nn &+ \eslash_1 \left(\frac{\pslash_t+\pslash_u}{2}+m \right)\eslash_2 - \eslash_2 \left(\frac{\pslash_t+\pslash_u}{2}+m \right)\eslash_1
    \Bigg] v_4.
\end{align}
Here, $p_t+p_u = p_3-p_4$ and $p_t-p_u = p_1 -p_2$.  Taking Eqs.~\eqref{eq:p3p4identity} and \eqref{eq:p1p2idensity} into account, we have
\begin{align}
    &\beta c_\theta (\Mt+\Mu) - (\Mt-\Mu)\nn
   =& \frac{2}{\hat s} \left[ \bar u_3 \eslash_1 v_4 (k-p_1)\cdot \epsilon_2 +   \bar u_3 \eslash_2 v_4 (p_2-k)\cdot \epsilon_1 + \bar u_3(\pslash_1-\pslash_2) v (\epsilon_1 \cdot \epsilon_2) \right],\nn
   =&2\Ms.
\end{align}
Therefore,
\begin{equation}\label{eq:MtupropMtu}
\frac{\Mt-\Mu}{2}+\Ms = \frac{\beta c_\theta}{2}(\Mt + \Mu),
\end{equation}
holds generally.  Substituting Eqs.~\eqref{eq:MtupropMtu},~\eqref{eq:AmpggttColorFactor}, and~\eqref{eq:colorsymasym} into Eq.~\eqref{eq:rhoqcd}, we find that the $\gamma\gamma \to t\bar t$ and $gg\to t\bar t$ processes produce exactly the same $t\bar t$ spin correlation.

\bibliographystyle{apsrev4-1}
\bibliography{refs}
\end{document}